\documentclass{article}
\usepackage{graphicx} 
\usepackage[margin=1in]{geometry}

\usepackage{amsmath, amssymb}
\usepackage{mathtools}
\usepackage{amsthm}
\usepackage{bbm}
\usepackage{enumitem}
\usepackage{natbib}
\usepackage{comment}
\usepackage{float}
\usepackage{hyperref}
\usepackage{setspace}
\usepackage{subcaption}
\usepackage{mathrsfs}
\usepackage{authblk}
\mathtoolsset{showonlyrefs}

\usepackage{xcolor}
\usepackage{tikz}
\definecolor{myblue}{RGB}{31,119,180}
\definecolor{myred}{RGB}{214,39,40}
\definecolor{mygreen}{RGB}{44,160,44}

\newtheorem{prediction}{Prediction}

\newtheorem{assumption}{Assumption}

\usepackage{setspace}

\usepackage[suppress]{color-edits}
\addauthor{wc}{red}
\addauthor{ga}{blue}
\addauthor{as}{orange}
\addauthor{nk}{magenta}
\addauthor{kz}{pink}
\newcommand{\email}[1]{\href{mailto:#1}{#1}}
\begin{document}
\title{Recommendation Quality and the Concentration of Consumption: Experimental Evidence from Netflix\thanks{Contact: Aridor: Netflix and Northwestern Kellogg, \email{garidor@netflix.com}. Chou: Netflix, \email{wchou@netflix.com}. Kallus: Netflix and Cornell University and Cornell Tech, \email{nkallus@netflix.com}. Scheid: Netflix, \email{ascheid@netflix.com}. Tran: Netflix, \email{atran@netflix.com}, Zielnicki: Netflix, \email{kzielnicki@netflix.com}. All authors were employed at Netflix during the duration of this project. We thank Rafael Jim\'enez-Dur\'an and Malika Korganbekova for helpful feedback and suggestions. We thank audiences at AI Algorithms and Markets Symposium (LUISS), Cornell Johnson, Conference on AI and Human Decisions (CMU), Netflix, and Research Roundtable on Platform Dynamics (Northwestern) for helpful comments. All errors are our own.}}
\author{Guy Aridor \qquad Winston Chou \qquad Nathan Kallus \\ Antoine Scheid \qquad Allen Tran \qquad Kevin Zielnicki}
\date{\today}
\maketitle

\thispagestyle{empty}
\begin{abstract}

We study an experiment with 8.5 million users on Netflix's recommender system to measure how improvements in recommendation technology affect the set of products that get consumed. Improvements increase total consumption and users' reliance on recommendations while diffusing recommendations and consumption away from the most popular products (``superstars") toward a larger number of moderately popular products (``middle-tail"), with minimal effects on the most niche products (``long-tail"). Our results challenge the notion that recommender systems polarize consumption -- raising the consumption shares of the head and tail at the expense of the middle -- and suggest that the returns to investing in middle-tail products grow as algorithms improve and platforms scale.

\end{abstract}
\newpage
\setcounter{page}{1}

\section{Introduction}

Recommender systems (RecSys), ubiquitous in the digital economy, use historical interaction data to connect users with relevant products. Their efficacy depends on extracting meaningful signals about consumer preferences from limited data. As this technology has improved, it has reshaped which products are consumed, with a large literature debating whether it concentrates consumption among the most popular products or spreads it across a long tail of niche products \citep{brynjolfsson2003consumer, salganik2006experimental, fleder2009blockbuster,  aridor2020deconstructing, calvano2025artificial}, with the products in the middle losing share to the extremes \citep{bar2012search}. 
However, the state-of-the-art algorithms have improved dramatically over the past several decades -- from item-based collaborative filtering algorithms \citep{sarwar2001item}, to matrix factorization techniques \citep{koren2009matrix}, to deep learning-based recommenders \citep{he2017neural}, and, more recently, to sequence and foundation models \citep{quadrana2018sequence, hsiao2025foundation}. These improvements continuously expand the frontier of goods that a RecSys can effectively target, yet the implications of these improvements for the concentration of consumption remain unknown.

This paper studies how improvements to recommendations reshape viewership using a unique experiment on Netflix's recommender system involving more than 8.5 million subscribers. Netflix's recommender system has been evolving with the progression in recommender systems research for the past twenty years, dating back to the Netflix Prize \citep{bennett2007netflix}, and drives approximately 80\% of the 97 billion hours of viewership on the platform in the first half of 2026 \citep{gomez2015netflix, netflix2026q2letter}. The experiment, a ``long-term holdback,'' compares a frozen algorithm with an evolving production treatment incorporating all algorithmic innovations deployed during a 60-day experimental period. Modern production recommendation systems are complex and so by focusing on a bundle of algorithmic changes to the production recommendation system, this experiment is insightful into how ongoing innovations in recommendation technology more broadly impact the concentration of consumption.

We find that ongoing improvements to the RecSys increase overall engagement on the platform and users' reliance on the recommendations for plays, primarily by shifting the recommendations -- and subsequently plays -- away from the most popular titles (``superstars") and toward moderately popular titles (``middle-tail"), with minimal effects on the long-tail. As recommendation technology matures, then, the plausible set of targetable titles continues to shift toward the middle-tail. We argue that these results are consistent with a stylized model in which algorithmic improvements allow the RecSys to better infer consumer preferences from a fixed amount of title-level interaction data. Since less popular titles generate fewer interactions, the model captures this data constraint as a tendency for the RecSys to default toward popular titles when preference signals for less popular titles are weak.\footnote{The microfoundation for this assumption comes from the literature in recommender systems documenting a \textit{popularity bias} -- an inherent tendency of deployed algorithms to favor popular titles because they learn from interaction data, which popular titles generate far more of \citep{celma2008hits, klimashevskaia2024survey}.} Improvements in recommendation technology attenuate this bias by allowing the RecSys to identify relevant titles with less data. The gains are largest for middle-tail titles: unlike superstars, they are not already reliably matched, and unlike long-tail titles, they generate enough data and command a large enough audience for better inference to matter. Improvements therefore shift recommendations toward these better-matched titles, raising overall engagement.

This paper contributes to a long-standing literature on how digitization has reshaped the types of goods that get consumed \citep{goldfarb2019digital, aguiar2024platforms}, and in particular to conflicting findings on whether recommendation systems concentrate or disperse consumption. Several early studies document that online consumption shifted away from popular titles \citep{brynjolfsson2003consumer, zentner2013video}, with a dramatic increase in the number of new products brought to market and substantial welfare gains \citep{waldfogel2017digitization, aguiar2018quality, reimers2021digitization, reimers2026ai}. Experimental evidence comparing personalized recommendations to non-personalized benchmarks similarly finds that personalization can increase engagement while dispersing aggregate consumption across products \citep{holtz2020engagement}. On the other hand, these technologies reduce search costs and thus concentrate consumption among superstars \citep{rosen_1981} and polarize the product distribution at the expense of middle-tail titles \citep{bar2012search}. Empirically, \cite{salganik2006experimental} show that popularity information can increase superstar concentration, while \cite{fleder2009blockbuster} and \cite{calvano2025artificial} show that popularity bias in classic RecSys algorithms can concentrate aggregate consumption on popular products.

These distributional effects are not fixed properties of recommendation systems. \cite{tucker2011does} show that popularity information reinforces superstar concentration under homogeneous preferences but can benefit niche products when preferences are heterogeneous. Our contribution is to identify a second, dynamic dimension -- algorithm maturity and the volume of title-level interaction data -- that changes which titles benefit as recommendation quality improves.\footnote{\cite{tucker2023algorithmic} highlights that algorithmic systems systematically underserve those for whom interaction data are sparse -- in our context, this manifests as a built-in disadvantage for less popular titles, consistent with broader evidence of diminishing returns to data \citep{peukert2024editor}. Our results show that this disadvantage is not permanent but diminishes as the algorithm matures.}  In our empirical context, \cite{zielnicki2026value} show using a structural model that previous RecSys improvements lift engagement by better identifying users with high incremental consumption from recommendations, and that this value is largest for middle-tail titles. We provide experimental evidence that continued improvements to the RecSys further shift consumption toward precisely this part of the distribution: away from superstars and toward middle-tail titles, while lifting engagement overall.

These findings have important implications for online platforms. By identifying which titles become more targetable as recommendation quality improves, our results speak to the frontier of products that benefit the most from algorithmic recommendation. In particular, continued improvements in recommendation technology mainly amplify products whose audiences can be reached through better targeting, but whose demand is difficult to realize under less mature systems or platforms with limited scale. This interpretation echoes supply-side evidence that digitization and streaming have increasingly supported middle-tail products: \cite{benner2023changing} document the sustainable production of medium-budget ``middle-tail'' films, while \cite{luke2025tales} finds that middle-tail books are overrepresented in online bookstores relative to brick-and-mortar retailers.

\section{Conceptual Model}\label{sec:model}

To tie together our empirical findings, we provide a stylized conceptual model of how improvements in recommendation technology change the concentration of consumption. The model is intended to capture the first-order economic forces at play not just on Netflix, but also in other markets where recommendation systems are prevalent, and to clarify the conditions under which our findings may generalize beyond our experimental setting.

\subsection{Setup}

We consider a representative user $i$ whose true utility from consuming title $j$ is given by:
\begin{align}
U_{ij} = \mu_{j} + \theta_{ij} ,
\end{align}
where $\mu_{j} \in \mathbb{R}$ is baseline popularity and $\theta_{ij}$ is drawn i.i.d. across $(i, j)$ with bounded, strictly positive density on $\mathbb{R}$ and finite mean. Thus, each title's utility is a combination of a vertical component ($\mu_j$) and a horizontal, idiosyncratic component ($\theta_{ij}$), so a randomly selected user would be more likely to prefer a more popular than a less popular title, but for some users a less popular title can provide higher utility.

We consider a stylized setup where there are three ``representative" titles -- long-tail (L), middle-tail (M), and superstar (S) -- and the user consumes at most one title.\footnote{While the treatment can, in principle, shift the popularity distribution, we validate in Appendix Figure~\ref{fig:bucket_stability} that the treatment in our empirical context is incremental enough that it does not dramatically change the title classification.} We assume that $\mu_{L} <\mu_M < \mu_S$ and that user $i$ has a title $j$ that provides the highest utility -- we denote this by $j^{*}_{i} = \arg \max_{j} U_{ij}$. Under this formulation, $p_{j} \; \coloneqq \;\Pr(j^{*}_{i} = j) \;\in\; (0,1)$ denotes the probability that title $j$ is user $i$'s best title. Thus, the goal of the RecSys is to identify this title.

\paragraph{User Choice.} The representative user receives a singleton recommendation $\mathcal{R}_{i} \subset \{ L, M, S\}$ and chooses from the three titles and an outside option. The user's decision utility $V_{ij}$ is influenced by the recommendation via a ``utility bonus" $b > 0$ that makes them more likely to consume the recommended title.  The bonus captures, in reduced form, the role of recommendations in reducing search frictions \citep{aridor2023economics, zielnicki2026value}. The user's decision utility for title $j$ is therefore given by:
\begin{align}\label{eq:V}
V_{ij} \coloneqq U_{ij} + b \cdot \mathbbm{1} \{ j \in \mathcal{R}_{i} \} ,
\end{align}
where $b >0$. The user chooses the title with the highest decision utility or the outside option (normalized to $V_{i0} = 0$), and we write the choice as $c_i = \text{argmax}_{j \in \{0, S, L, M\}} V_{ij}$.\footnote{We assume that, as our experiment is conducted over 60 days, the experimental intervention only changes \textit{which} titles enter into $\mathcal{R}_i$ and not how users internalize the signal provided by the recommendations.  Thus, the value of $b$ remains constant throughout the experiment. However, all of our predictions would similarly hold as long as $b$ increases due to the improved recommender.}

\paragraph{Platform Recommendation Technology.} The platform's recommendation technology is indexed by a single parameter $\delta$ and we make the following assumptions about it.

\begin{assumption}[Recommendation Technology]\label{a:tech}
The RecSys attempts to identify the user's best title with a classifier whose accuracy is summarized by the index $\delta$. We assume:
\begin{itemize}
    \item \textbf{Popularity default.} Conditional on $j^{*}_{i} = j$ and the user's full utility, the recommender identifies $j$ with probability $q_{j}(\delta)$ and otherwise outputs $S$ (i.e., for user $i, \Pr(\mathcal R_i = \{j\} \mid U_i, j_i^\star=j) = q_j(\delta)$).
    \item \textbf{Monotonicity.} The functions $q_L, q_M$ are differentiable and weakly increasing in $\delta$.
    \item \textbf{Middle-tail dominance.} Over the empirically relevant range of $\delta$, the marginal precision gain is largest for the middle-tail:
    $p_M\, q_M'(\delta) \;>\; p_L\, q_L'(\delta) \;\geq\; 0$.
\end{itemize}
\end{assumption}

The primary justification for this assumption comes from the \textit{popularity
bias} widely documented in the recommendation systems literature \citep{celma2008hits,
klimashevskaia2024survey}, by which deployed algorithms over-recommend popular titles because less popular titles generate too little data to extract a reliable signal. Appendix~\ref{sec:recsys_microfoundation} motivates the assumption with a Bayesian recommender that generates this bias endogenously and shows the popularity bias and, over the relevant range of $\delta$, that the weighted identification gain $p_j q_j'(\delta)$ is largest for the middle-tail: the superstar's $q_S'(\delta)$ is small, since it is already well targeted, and while the long-tail's $q_L'(\delta)$ may be sizeable, too few users rank it first for the identification gain to matter. One important implication from the Bayesian model is that the middle-tail dominance is not a universal feature across all $\delta$, but rather only holds for most intermediate values of $\delta$ where users who prefer $S$ are always matched and there are still gains from matching users who prefer $M$ to their preferred title, which we believe is reasonable in our context given the maturity of the Netflix RecSys.

\paragraph{Generalizability.} The model has several key ingredients for understanding how improved recommendation technology influences choices that are not just useful for rationalizing behavior on Netflix, but also in other markets where recommendation systems are prevalent and users consume goods without additional monetary payment, such as music, books, and social media. First, there is both horizontal and vertical differentiation, so that there is scope for personalization by finding better horizontal matches as well as a ``fallback'' set of broadly popular titles when information about these matches is limited. The latter is consistent with the large popularity bias literature in recommender systems \citep{celma2008hits, klimashevskaia2024survey}. Second, we expect this to generalize to other markets where recommendations play a large role in driving choices, particularly where there is a large set of possible choices and users face search frictions. Third, the prediction that technological improvements primarily benefit middle-tail goods applies where popular goods are already reliably matched, middle-tail goods generate enough interaction data for improved technology to identify better horizontal matches, and long-tail goods still generate too little data for these improvements to have a substantial effect. Given the maturity of the Netflix RecSys, we expect that our empirical setting provides insight into markets where these conditions hold.

\subsection{Empirical Predictions}

Our experimental intervention involves an increase in the effectiveness of recommendation technology ($\delta$) and we characterize directional predictions from this change, holding all other parameters fixed. Each prediction relies on Assumption~\ref{a:tech} with proofs deferred to Appendix~\ref{sec:proofs}.\footnote{The empirical measures and predictions are those that are derived from our theory -- they are not the core product metrics that Netflix optimizes for.}

The first natural consequence of the improvement in recommendation technology is that it shifts recommendation share away from the popularity default and to the tail of the distribution. We define the recommendation probability $f_j(\delta) \coloneqq \Pr(j \in \mathcal{R}_i(\delta))$ and characterize how it changes:

\begin{prediction}[Hump-shaped recommendations]
\label{prop:hump}
An increase in $\delta$ results in a hump-shaped redistribution of recommendation share with the largest gains for the middle-tail (M):
$f_M'(\delta) \;>\; f_L'(\delta) \; > \;  f_S'(\delta)$.
\end{prediction}

Intuitively, Prediction~\ref{prop:hump} follows since some users prefer $M$ to $S$, but the platform observes too noisy of a signal for $\theta_{iM}$ in order to recommend $M$ over $S$. Increasing $\delta$ provides additional precision for the platform's estimated score for title $M$, leading to an increased fraction of users who prefer $M$ to receive it in the recommendations instead of $S$. A similar pattern holds for users who prefer $L$, but the magnitude is smaller since the popularity gap between $L$ and $S$ is bigger and the signal is still noisy even after the technology improvement.

If recommendations change (Prediction~\ref{prop:hump}) and recommendations matter for choice ($b > 0$), then naturally consumption should shift to the newly recommended titles. We denote $\pi_j(\delta) \coloneqq \Pr(c_i = j)$ as the consumption probability for title $j$ and characterize how it changes across the three titles:
  \begin{prediction}[Play Share Rotation]\label{prop:shares}
An increase in $\delta$ shifts consumption from the superstars to the middle-tail and long-tail:  $\pi_M'(\delta) > 0$ and $\pi_L'(\delta) \geq 0$, while $\pi_S'(\delta) < 0$.
\end{prediction}

Predictions~\ref{prop:hump} and~\ref{prop:shares} establish that recommendations and consumption shift toward the middle-tail, but redistribution alone does not reveal whether users are better served: an algorithm could diversify recommendations across the middle-tail without those being titles users want to consume \citep{chen2024impact}. The key test is whether the redistribution raises overall \emph{engagement} -- the likelihood that a user does not take the outside option -- since users consume the newly recommended titles only if they prefer them to the outside option:
\begin{prediction}[Engagement Levels]\label{prop:engagement}
    An increase in $\delta$ increases total engagement: $\Pr(c_i \neq 0)$ is strictly increasing in $\delta$.
\end{prediction}
The main force behind this prediction is that some users who preferred $M$ or $L$ over $S$ were previously recommended $S$ due to the popularity bias, which led them to consume nothing; as the RecSys shifts to their preferred title, they consume it and the outside-option share falls. Because this newly realized consumption comes from the recommended title, the share of consumption originating from recommendations should rise as well:
\begin{prediction}[Share of Plays from Recommendation]\label{prop:recdep}
    An increase in $\delta$ results in a larger share of consumption originating from recommendations: $\Pr(c_i \in \mathcal{R}_i \mid c_i \neq 0)$ is strictly increasing in $\delta$.
\end{prediction}

Finally, it is important to understand whether, conditional on consumption, the improved recommendations lead users to better-matched titles. This is theoretically ambiguous in our model as there are two counteracting forces: \textit{inframarginal switching} may increase average match quality if the improved recommendations shift a user's consumption from $S$ to their preferred title, but \textit{extensive-margin entry} may decrease average match quality by inducing users who would otherwise consume the outside option to consume marginal titles. This leads to our final prediction (which is formally stated in Appendix~\ref{sec:proofs}), which ultimately will be an empirical question to understand which force wins out:

\begin{prediction}[Match Quality]\label{prop:match_quality}
An increase in $\delta$ results in weakly better average match quality if and only if the match-quality gains from inframarginal switching weakly exceed the composition effect from extensive-margin entry. 
\end{prediction}

\section{Data and Field Experiment}

\begin{figure}[ht]
\centering
\caption{Netflix Homepage}
\label{fig:netflix_homepage}
\includegraphics[width=0.6\textwidth]{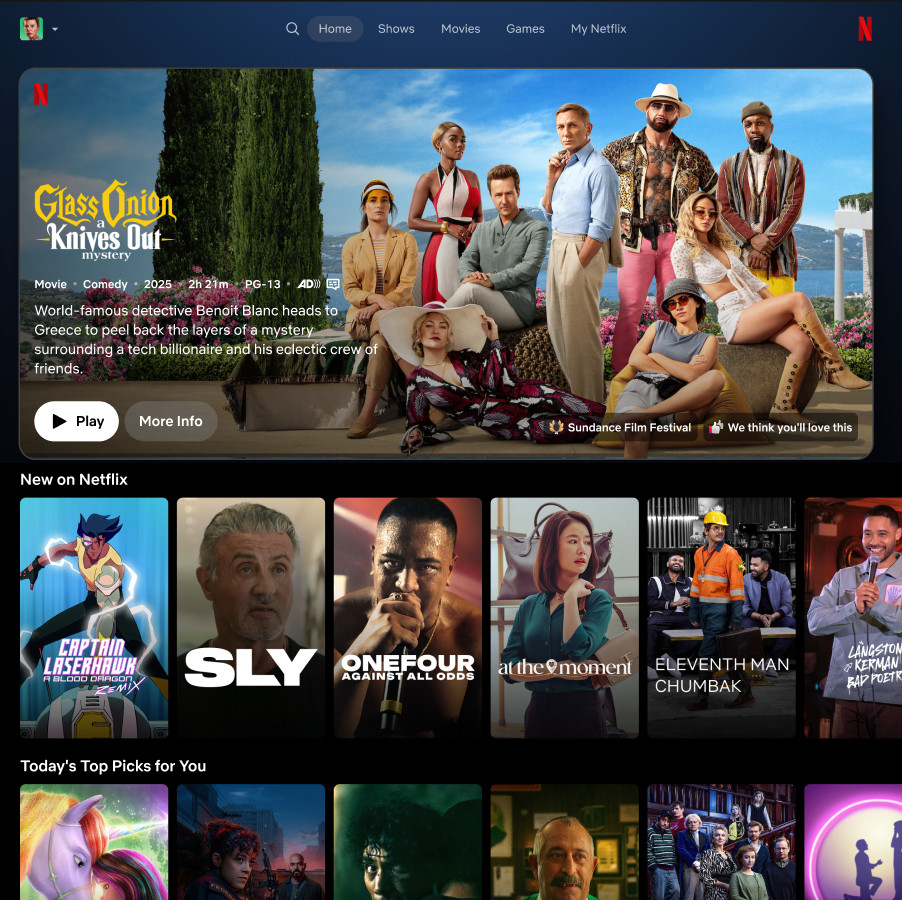}
\end{figure}

\noindent \textbf{Data and Empirical Setting.} Our empirical context is the video-streaming platform Netflix. Users pay a subscription fee to access Netflix and can then costlessly play any title on the platform. We therefore operationalize consumption in our empirical context as playing a title, and define engagement as the overall volume of consumption. The set of titles consists of television shows (which have multiple episodes and seasons) and movies. The platform's homepage -- which is personalized to each user -- is shown in Figure \ref{fig:netflix_homepage}. Users can directly play titles from the recommendations (defined as titles on the homepage), continue watching titles they had started before, or manually search for titles using the search bar. Our analysis uses platform data on plays and recommendations for each user in the experiment.

\noindent \textbf{Field Experiment.} To evaluate the empirical implications of our model, we analyze an experiment on Netflix on 8,559,252 subscribers over a 60-day period from February to April 2025.  This experiment was designed as a ``cumulative holdback'' test to measure the collective impact of multiple innovations to personalization algorithms during this period. The control arm of the experiment corresponds to the state of the recommendation system prior to the start of the test, whereas the treatment arm corresponds to the ``production'' state of Netflix that is experienced by the majority of Netflix members and continuously updated during the test as new innovations (e.g., algorithmic updates) are released. These releases are held back from the control group, which continues to experience the pre-experiment version of the recommendation system. Such holdback tests are run periodically on a small fraction of the Netflix member base in order to measure long-term progress on personalization \citep{kohavi2020trustworthy}.

Together with the maturity of the Netflix RecSys and the important role of recommendations in driving consumption on the platform, this makes the experiment informative about improvements to mature recommendation systems in markets where the conditions described in Section~\ref{sec:model} hold. First, the treatment includes all innovations made to personalization algorithms during the experiment timeframe. As such, it is broadly representative of ongoing efforts to improve personalized recommendations at Netflix. Second, Netflix's recommender system has continually been on the cutting edge of recommender system innovation for two decades, dating back to the Netflix Prize \citep{bennett2007netflix}, which provides a meaningful benchmark to understand how improvements to recommendation technology at the current frontier impact the concentration of consumption beyond Netflix. Third, the Netflix RecSys is an important driver of consumption on the platform, with 80\% of the nearly 97 billion view hours sourced from the recommendations \citep{gomez2015netflix, netflix2026q2letter}. Finally, each change was individually tested to ascertain improvements in isolation and the relatively long duration of the experiment helps us confirm that the following impacts are stable and persistent.

The innovations deployed during the experiment are to the existing recommendation system stack in 2025. As has been publicly shared, the Netflix recommender system traditionally operates as a multi-stage process wherein a page-generation algorithm selects and orders rows and row-specific rankers order titles within rows \citep{wang2026genpage}. The treatments bundled into this holdback fall into four categories of routine technical changes that impact both stages: (i) adjusting model architecture and training procedure for the ranker, such as moving to a multi-task model with several prediction heads \citep{oh2024intentrec}; (ii) reweighting the user actions used to weight training examples, such as plays, play duration, and thumbs in the reward function \citep{tang2023reward, netflix2024longterm}; (iii) introducing specialized rankers for specific user interfaces within the product; and (iv) updating the features, embeddings, and label-generation pipelines used for model training.

Overall, these treatments represent routine changes within the existing recommendation stack aligned with the structure of the model in Section~\ref{sec:model} and Appendix \ref{sec:recsys_microfoundation} -- changes in the training pipelines and modeling architecture to extract more precise signals from the existing interaction data. To the extent that these changes improve the system’s ability to identify relevant titles, they serve as an increase in $\delta$ in the language of our model. Crucially, no treatment was designed to target titles by popularity, and none optimized specifically for plays of middle-tail titles. More broadly, the Netflix recommender optimizes for user engagement signals, such as plays and thumbs, that serve as proxies for long-term member retention \citep{tang2023reward, netflix2024longterm}.\footnote{\url{https://research.netflix.com/research-area/recommendations} provides more information on historical changes to the Netflix RecSys that have been made public.} Given that these interventions primarily sought to improve user engagement with recommendations, rather than lift plays of middle-tail titles, they constitute an appropriate test of our model without mechanically ensuring our empirical findings on the concentration of consumption. 

%

\section{Distributional Shifts in Recommendations and Plays}\label{sec:distributional_shifts}

In this section we characterize the \textit{distributional} shifts as a result of the technology increase by first exploring its impact on the distribution of recommendations (Section \ref{subsec:recommendations_distributional_shift}) and then its subsequent impact on the distribution of played titles (Section \ref{subsec:consumption_distribution_shift}).

A key challenge is measuring distributional impacts without letting our intervention contaminate the play distribution itself. Unfortunately, as the environment is dynamic with many titles entering and exiting the platform, pre-experimental shares are not representative of shares during the experiment. As such, we partition titles into the same buckets as in the conceptual model and assign them based on their percentile on the aggregate play distribution across the treatment and control groups: ``long-tail" (L), ``middle-tail" (M), or ``superstar" (S) buckets -- where ``superstar" is defined as the top 5 percentile, the ``long-tail" as the bottom 50 percentile, and the ``middle-tail" as the set of titles in between.\footnote{While the treatment can influence the empirical popularity distribution, we empirically confirm in Figure~\ref{fig:bucket_stability} that these buckets remain similar between our treatment and control groups.}

Our primary specification measures how treatment reallocates a user's recommendations and plays across title buckets. Let $c^{o}_{ik}$ be the total number of instances for outcome $o$ for a given user $i$ and titles in bucket $k$. The outcome of interest is the within-user share $s^{o}_{ik} \;=\; c^{o}_{ik} / \sum_{k'} c^{o}_{ik'}$ for users with at least one outcome. We estimate, separately for each outcome $o$ and each bucket $k$:
\begin{equation}\label{eq:across_groups}
s^{o}_{ik} \;=\; \alpha^{o}_{k} \;+\; \beta^{o}_{k}\, T_{i} \;+\; \varepsilon^{o}_{ik},
\end{equation}
\noindent
where $T_i$ is an indicator for user $i$ being assigned to the treatment arm, $\alpha^{o}_{k}$ is the mean share in the control and $\beta^{o}_{k}$ is the average treatment effect on that share. We compute robust standard errors.

\subsection{Recommendations Shift to the Middle-Tail}\label{subsec:recommendations_distributional_shift}

\begin{figure}[ht]
    \centering
    \caption{Treatment Effects on Recommendation Shares}
    \label{fig:rec-shares}
    \begin{subfigure}[t]{0.30\linewidth}
        \centering
        \includegraphics[height=2in]{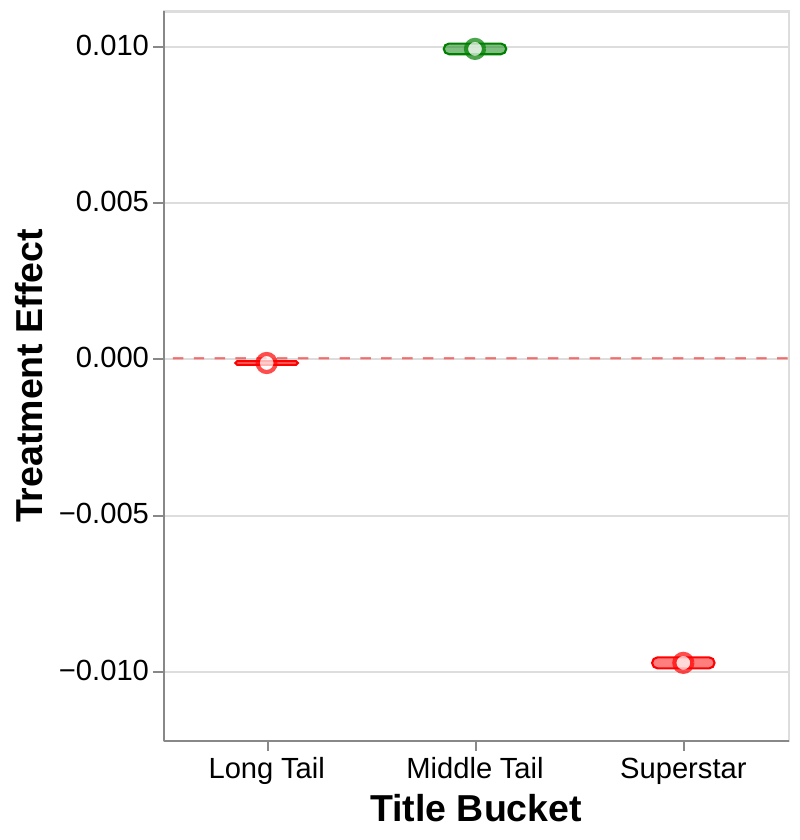}
        \caption{Shares by Bucket}
        \label{fig:rec-shares-avg}
    \end{subfigure}
    \hfill
    \begin{subfigure}[t]{0.52\linewidth}
        \centering
        \includegraphics[height=2in]{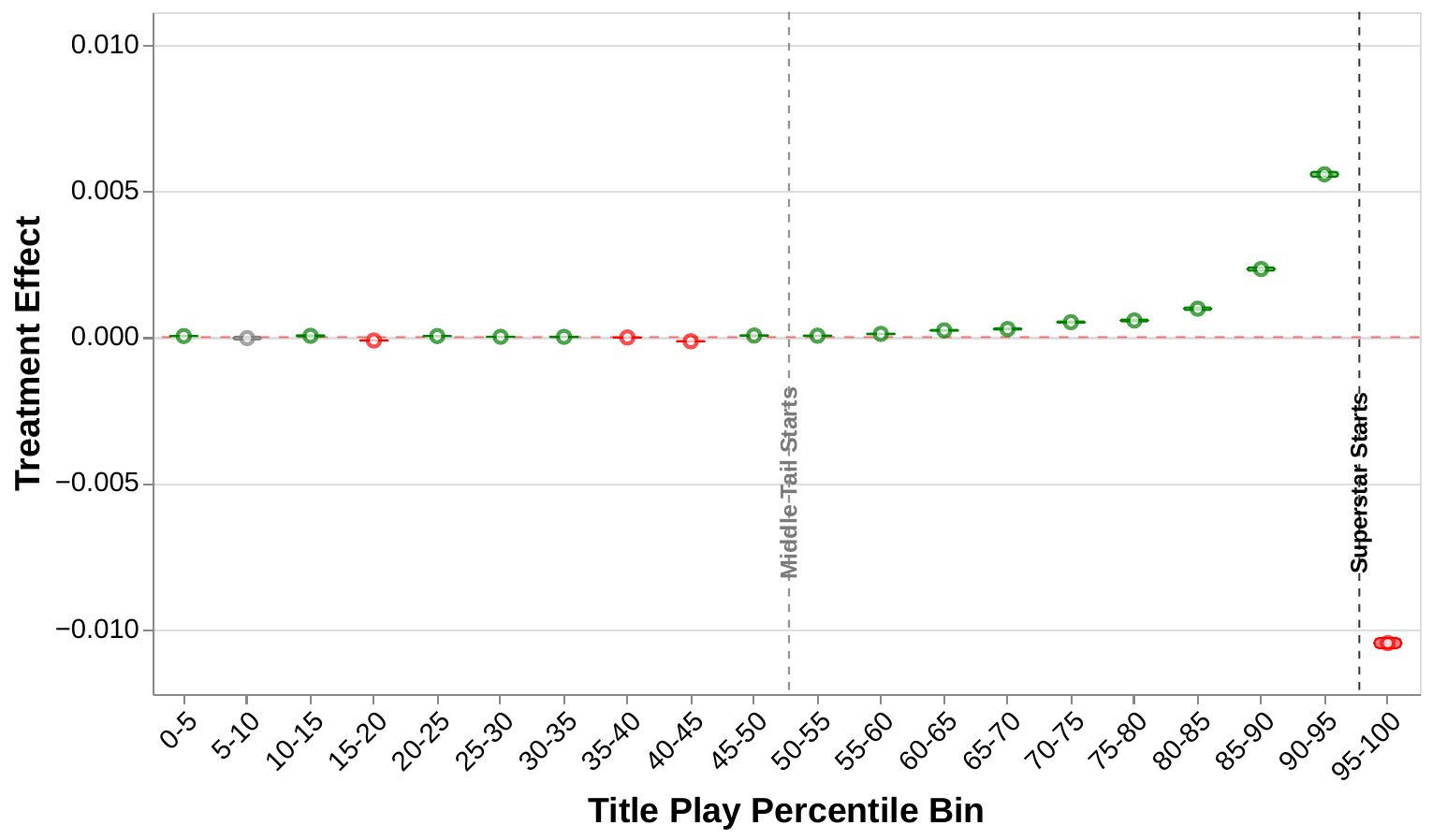}
        \caption{Shares by Ventile Bins}
        \label{fig:rec-shares-ventiles}
    \end{subfigure}
    \hfill
    \begin{subfigure}[t]{0.14\linewidth}
        \centering
        \includegraphics[height=2in]{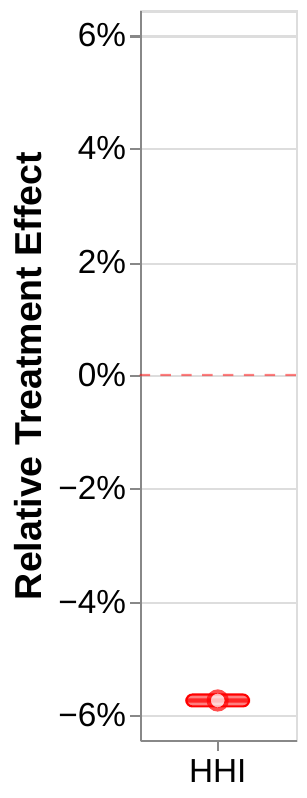}
        \caption{HHI}
        \label{fig:rec-shares-hhi}
    \end{subfigure}
 
    \caption*{\footnotesize \textsc{Notes}: Estimated Average Treatment Effects (ATEs) on the distribution of recommendations. Panel (a) presents coefficient estimates from~\eqref{eq:across_groups} with the dependent variable of the share of an individual's recommendations accounted for by each bucket. Panel (b) presents  treatment effects on recommendation share within each pooled ventile, where the ventiles are defined using the pooled play distribution across the treatment and control groups. Panel (c) presents coefficient estimates for the changes in the recommendations HHI, where we compute standard errors using a delete-one jackknife. Bars denote 95\% confidence intervals.}
\end{figure}

The first implication of an increase in the efficacy of recommendation technology, as stated in Prediction~\ref{prop:hump}, is that recommendations redistribute away from superstars toward the middle-tail, with minimal effects on the long-tail. We test this by estimating specification~\eqref{eq:across_groups} on the recommendations users observe during the experiment for each of the title buckets $\{ L, M, S \}$.

Figure~\ref{fig:rec-shares-avg} confirms the hump shape predicted by Prediction~\ref{prop:hump}: recommendation share shifts away from superstar titles and primarily gets redirected to middle-tail titles, with quantitatively small changes to the long-tail. Figure~\ref{fig:rec-shares-ventiles} shows that this same pattern persists using more granular ventile buckets and Appendix Figure~\ref{fig:rec-shares-control-only} shows that these patterns replicate when we take the most conservative approach of defining the percentiles purely in terms of the play percentile in the control group.\footnote{Appendix Figure~\ref{fig:recommendation_quantile_treatment_effects} additionally shows that within the superstars the effect is most stark for the highest percentile.}

An important consequence of this redistribution is that it shifts recommendation share from a small number of superstar titles and disperses it to a large number of middle-tail titles that are better fits for users. As such, the shift to the middle-tail should decrease the amount of concentration in recommendations across users. We estimate the effects of the technology improvement on the Herfindahl–Hirschman Index ($\mathrm{HHI}= \sum_{j=1}^{N} s_j^2$) -- a standard measure for the degree of concentration that we compute at the title level -- to test this. Figure~\ref{fig:rec-shares-hhi} confirms that HHI has a statistically and economically significant decrease of 5.7\%.

\subsection{Plays Shift to the Middle-Tail}\label{subsec:consumption_distribution_shift}
\begin{figure}[ht]
    \centering
    \caption{Treatment Effects on Play Shares}
    \label{fig:consumption-shares}
    \begin{subfigure}[t]{0.30\linewidth}
        \centering
        \includegraphics[height=2in]{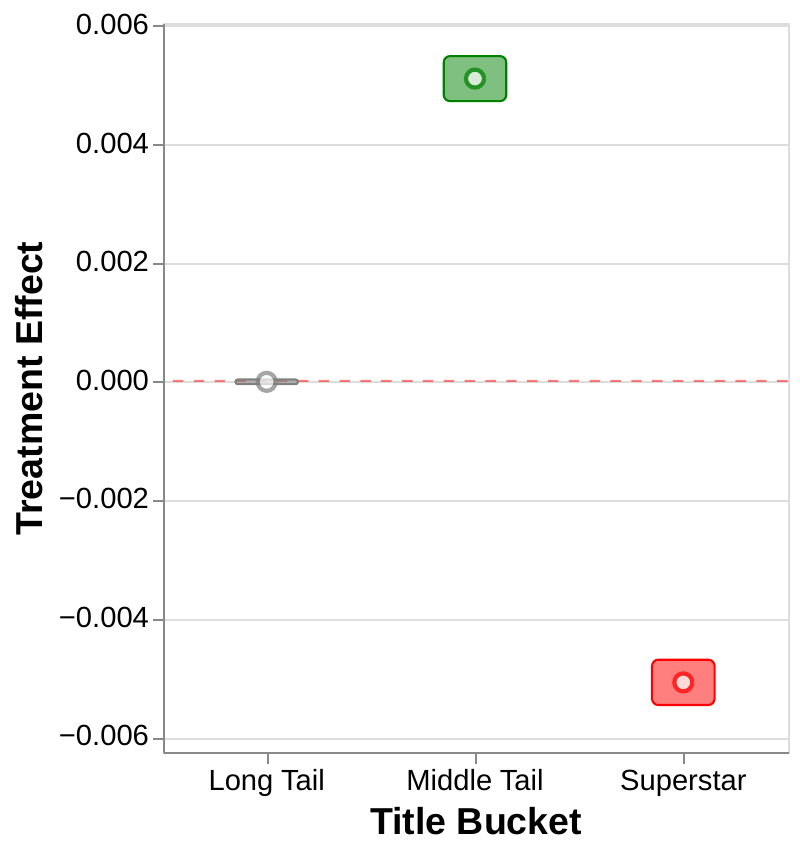}
        \caption{Shares by Bucket}
        \label{fig:consumption-shares-avg}
    \end{subfigure}
    \hfill
    \begin{subfigure}[t]{0.52\linewidth}
        \centering
        \includegraphics[height=2in]{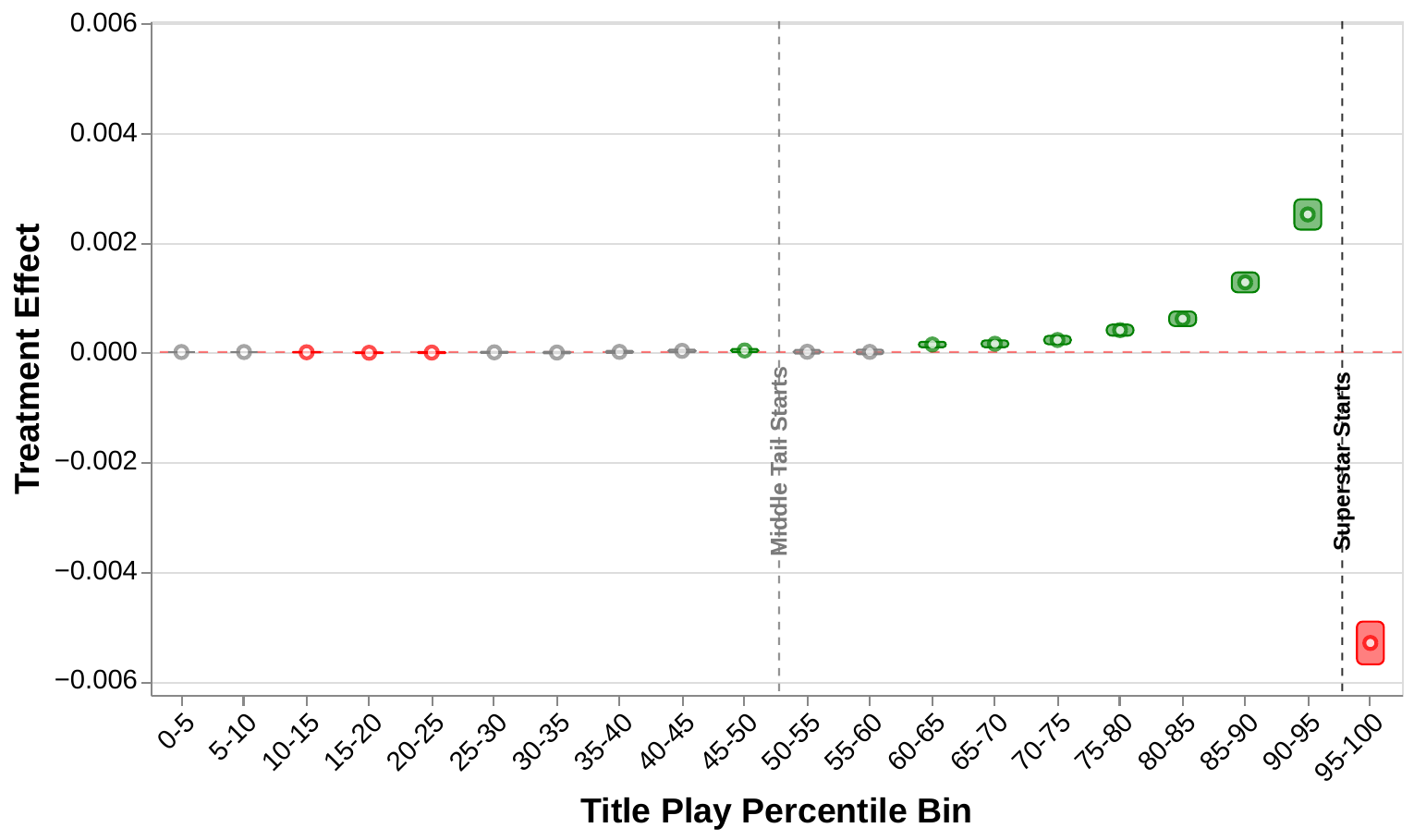}
        \caption{Shares by Ventile Bins}
        \label{fig:consumption-shares-ventiles}
    \end{subfigure}
    \hfill
    \begin{subfigure}[t]{0.14\linewidth}
        \centering
        \includegraphics[height=2in]{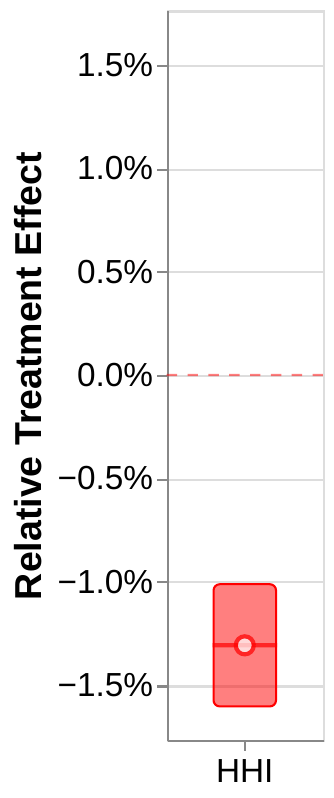}
        \caption{HHI}
        \label{fig:consumption-shares-hhi}
    \end{subfigure}
 
    \caption*{\footnotesize \textsc{Notes}: Estimated ATEs on the distribution of title plays during the experiment. Panel (a) presents coefficient estimates from specification~\eqref{eq:across_groups} with the dependent variable of the share of an individual's plays accounted for by each bucket. Panel (b) presents treatment effects on play share for each pooled ventile, where ventiles are defined using the pooled play distribution across the treatment and control groups. Panel (c) presents coefficient estimates for the changes in the Herfindahl-Hirschman Index of plays, where we compute standard errors using a delete-one jackknife. Bars denote 95\% confidence intervals.}
\end{figure}

A natural implication of the shift in the distribution of recommendations is that it may also shift the distribution of plays. As shown in our conceptual model, plays depend on the title's recommendation probability \textit{and} the probability that the user follows the recommendation. If the RecSys diversified the recommendations by replacing superstar titles with middle-tail titles that were not appealing, then users could still seek out the superstars via search or decide not to play anything on the platform at all, leading to minimal differences in the play share. Prediction~\ref{prop:shares} indicates that play share shifts if the recommendations are sufficiently consequential in user choices (i.e., $b > 0$).

We test Prediction~\ref{prop:shares} by estimating specification~\eqref{eq:across_groups} for each title bucket, reporting the estimates in Figure~\ref{fig:consumption-shares-avg}. The results confirm the prediction: superstar play share falls with most of the redistribution going to the middle-tail, with the long-tail quantitatively unchanged. Figure~\ref{fig:consumption-shares-ventiles} shows that this same pattern persists using more granular ventile buckets and Appendix Figure~\ref{fig:consumption-shares-control-only} shows that these patterns replicate when we take the most conservative approach of defining the percentiles purely in terms of the play percentile in the control group.\footnote{Appendix Figure~\ref{fig:consumption_quantile_treatment_effects} additionally shows that within the superstars the effect is most stark for the highest percentile.}

Finally, we measure whether this corresponds to a reduction in play concentration: Figure~\ref{fig:consumption-shares-hhi} shows that HHI falls by a statistically and economically significant 1.2\%, consistent with viewership dispersing from superstar titles to more personalized middle-tail titles.

\section{Engagement Levels and Match Quality}\label{sec:overall_engagement}

While the results in Section \ref{sec:distributional_shifts} document that both recommendations and plays shift toward the middle-tail and subsequently become less concentrated, the key test of whether this shift is a ``technological improvement" is if these distributional shifts translate into a corresponding improvement in the consumer experience that is driven by the recommendations. We assess this in this section by first characterizing whether and why the improvement leads to an increase in the \textit{overall level} of engagement (Section \ref{subsec:overall_consumption}) and then assessing whether the played titles had similar match quality (Section \ref{subsec:match_quality}).

For most of this analysis, we compare \textit{user-level} outcomes across the experimental period and estimate average treatment effects using the following regression:
\begin{equation}\label{eq:ate_reg}
Y_{i} = \alpha + \beta \, T_i + \varepsilon_{i} ,
\end{equation}
with similar notation as before and compute robust standard errors.

\subsection{Overall Engagement and Recommendation Dependence}\label{subsec:overall_consumption}

The simple conceptual model from Section~\ref{sec:model} predicts that the recommendations should be more relevant to consumers, inducing increased engagement (Prediction~\ref{prop:engagement}) that is driven by increased play share coming from the recommendations (Prediction~\ref{prop:recdep}). In this section, we test these predictions.

\begin{figure}[ht]
    \centering
    \caption{Treatment Effects on Overall Engagement}
    \label{fig:overall-engagement}
        \centering
        \includegraphics[width=0.8\linewidth]{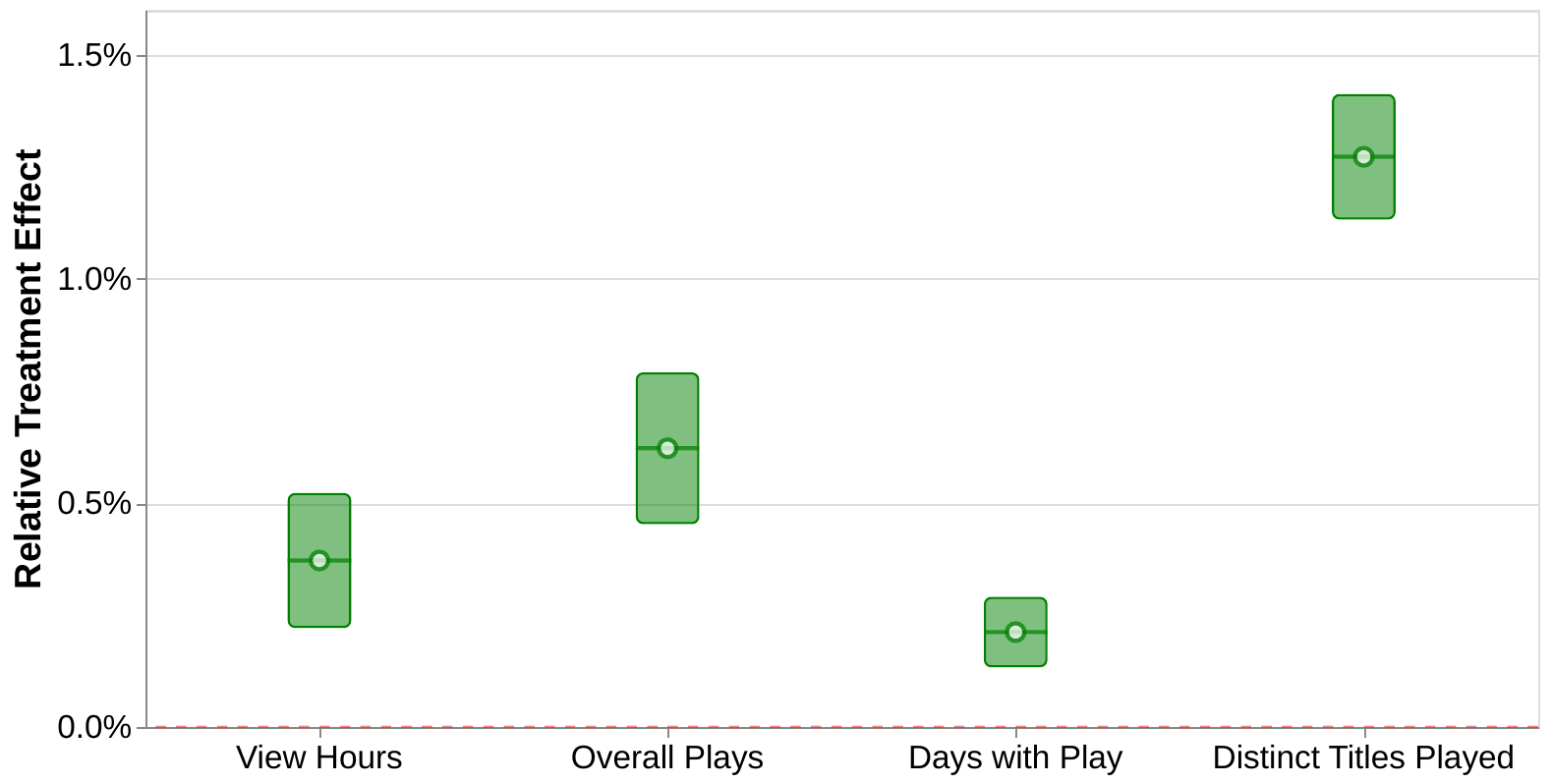}
        \caption*{\footnotesize \textsc{Notes}: Estimated ATEs from~\eqref{eq:ate_reg} on view hours, overall plays, days with at least one play, and distinct titles played. Effects are reported as relative percent change against the control mean. Bars denote 95\% confidence intervals computed via robust standard errors.}
    \vspace{1em}
\end{figure}

Figure~\ref{fig:overall-engagement} presents the estimates of $\beta$ from \eqref{eq:ate_reg} across four different measures of engagement: total view hours, number of plays, the number of days with at least one play, and the number of distinct titles played. The results confirm Prediction~\ref{prop:engagement}: view hours increased by 0.37\%, the total number of titles played by 0.62\%, the number of distinct titles played played by 1.2\%, and the number of days with at least one play by 0.21\%.\footnote{For television shows, total and distinct title counts can diverge, since each episode is counted separately toward the total but does not count as a distinct title. Appendix Figure~\ref{fig:engagement_title_type} shows that the increase in engagement is for both TV shows and movies, but leads to larger increases for TV shows.}$^{,}$\footnote{Appendix Figure~\ref{fig:over_time_treatment_effects} shows that the treatment effects are stable or increasing, indicating sustained improvements in engagement throughout the intervention period.} As context for the magnitudes, \cite{zielnicki2026value} estimate via a structural model that reverting to the predominant RecSys algorithm from 10 years ago (matrix factorization) would reduce days with a play by 4\%, with economic effects comparable to those of reverting from matrix factorization to popularity-based ranking (then valued at more than \$1 billion by \cite{gomez2015netflix}). Our estimate of the effects of roughly two months of algorithmic innovation -- a 0.21\% increase in days with a play -- represents about 5\% of this magnitude.  Crucially, it implies that the shift away from superstars documented in Section~\ref{sec:distributional_shifts} is not diversification for diversification's sake, but actually coincides with improved engagement.

The intuition from our conceptual model for \textit{why} the improvement leads to an increase in overall engagement is simple: the RecSys replaces less well-matched superstars with better matched middle-tail titles which results in users being more likely to find content to play from the recommender system. In our empirical context, we can validate this by measuring whether the reliance on the recommendations for generating plays increases -- which is the crux of Prediction~\ref{prop:recdep}. 

\begin{figure}[ht]
    \caption{Treatment Effects on Reliance on Recommendations}
    \label{fig:rec_dependence}
    \begin{center}
\includegraphics[width=0.8\textwidth]{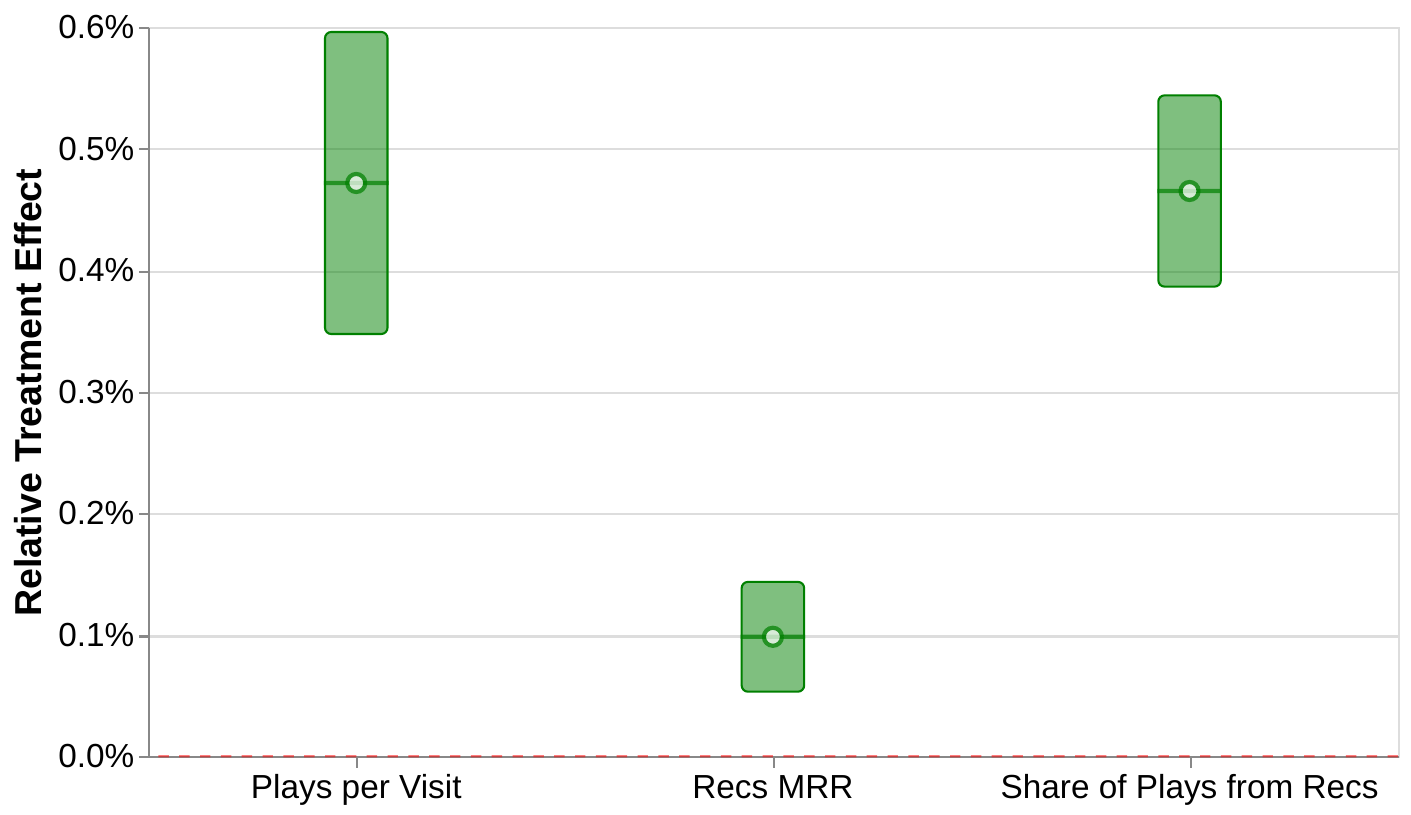}
\caption*{\footnotesize \textsc{Notes}: Estimated ATEs from~\eqref{eq:ate_reg} for measures of reliance on recommendations: plays per visit, mean reciprocal rank (MRR), and the share of first-in-session plays sourced from recommendations rather than other navigation. MRR is computed using the reciprocal of the row at which the played title appears on the homepage. All effects are reported as relative percent change against the control mean. Bars denote 95\% confidence intervals computed using robust standard errors.}
\end{center}
\end{figure}

To test Prediction~\ref{prop:recdep}, we restrict focus to the first play per session in order to rule out effects coming from continuation play and classify each play as either coming from recommendations, manual search, play continuation, or previously saved (e.g., my list). We then estimate~\eqref{eq:ate_reg} for the ratio of plays to visits, the share of plays that originate from recommendations, and the mean reciprocal rank (MRR) of plays.\footnote{MRR measures the reciprocal of the row number on the homepage from which the play was derived and is a common measure used to assess recommendation quality \citep{gunawardana2009survey} with a larger value indicating that the played title was ranked higher in the recommendations.} Figure~\ref{fig:rec_dependence} shows that the ratio of plays to visits increases by 0.47\%, the share of plays originating from the recommendations increases by 0.5\%, and MRR increases by 0.1\% -- all consistent with Prediction~\ref{prop:recdep}.

\subsection{Match Quality}\label{subsec:match_quality}

We now evaluate Prediction~\ref{prop:match_quality} -- how match quality changes due to the treatment. We rely on two proxies for match quality. The first is whether users continue watching the titles they start. As titles have different runtimes, we denote a title as \textit{completed} if the user watched it for at least $\min \{X, \text{runtime}\}$ minutes for varying values of $X$ on the first day of viewing the title. The second proxy is whether a user provided a positive ``thumbs'' rating to a title. We restrict ourselves to titles for which the user was shown the thumbs prompt, which asks the user for a negative thumb (thumbs down) or a positive thumb (single or double thumbs up).

\begin{figure}[ht]
    \centering
        \caption{Treatment Effects on Match Quality}\label{fig:match-quality}
    \begin{subfigure}{0.34\linewidth}
        \centering
        \includegraphics[height=1.2in]{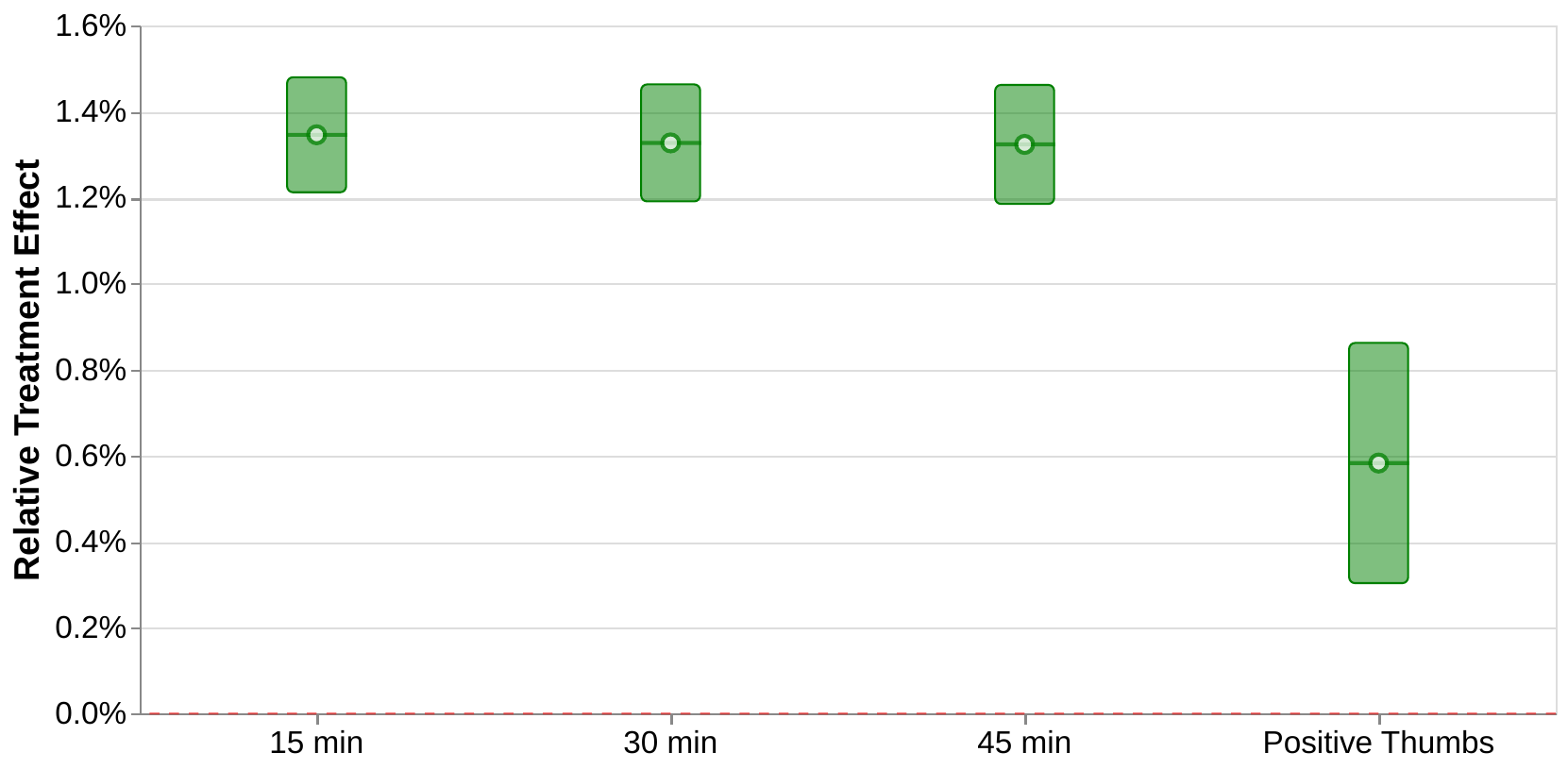}
        \caption{Total Number of Played Titles}\label{fig:match_quality_levels}
    \end{subfigure}
    \hfill 
    \begin{subfigure}{0.64\linewidth}
        \centering
        \includegraphics[height=1.2in]{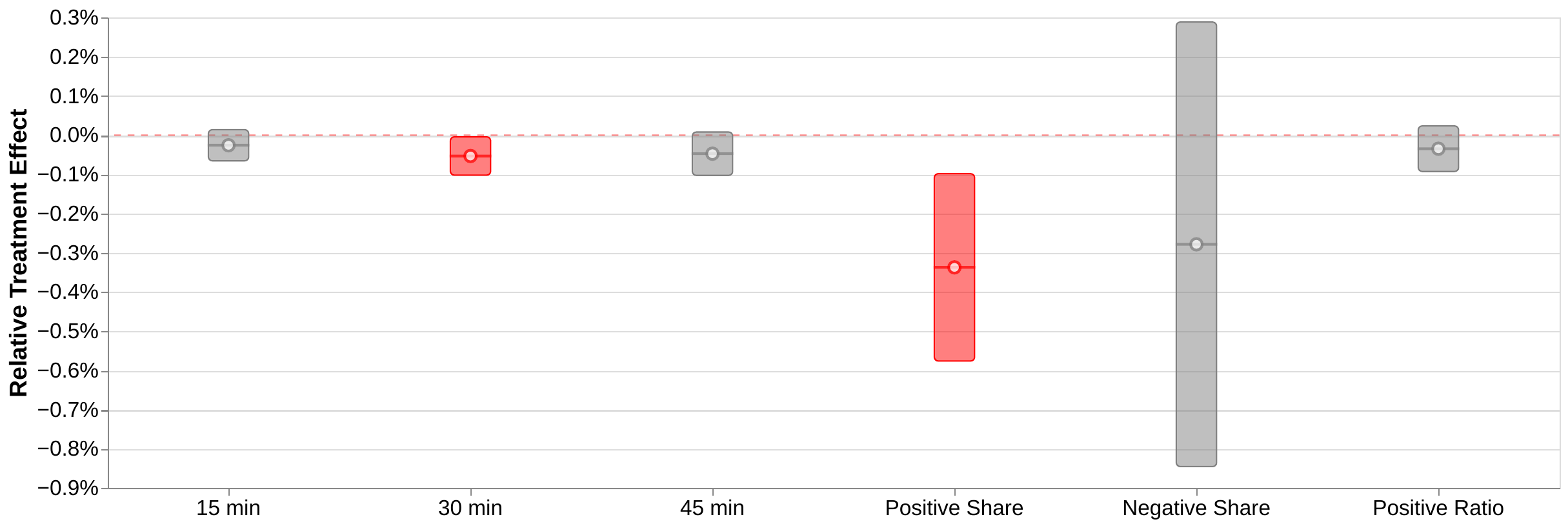}
        \caption{Fraction of Played Titles}\label{fig:match_quality_frac}
    \end{subfigure}
   \caption*{\footnotesize \textsc{Notes}: Estimated ATEs from \eqref{eq:ate_reg} on match-quality measures: shares of titles thumbed positively/negatively, and shares watched at least 15, 30, or 45 minutes (or full runtime, if shorter) on the first viewing day. Watch-duration shares use all plays; thumbs shares are restricted to plays long enough to trigger the thumbs prompt. Panel (a) shows estimates for the total titles per user reaching the watch-time threshold or receiving a positive thumbs; Panel (b) shows the user-level fraction of titles meeting each duration, receiving a positive/negative thumbs, or the positive-to-negative thumbs ratio. Bars are 95\% CIs from robust standard errors.}
\end{figure}

First, we measure whether users find more titles that they complete or thumb positively. We estimate~\eqref{eq:ate_reg} on the number of titles satisfying each of these measures and report the results in Figure~\ref{fig:match_quality_levels}: users complete 1.3\% more titles at each of the $X = \{15, 30, 45\}$-minute thresholds and provide 0.57\% more positive thumbs.

A natural question -- and the crux of Prediction~\ref{prop:match_quality} -- is whether the average title that a user plays is of comparable match quality. For each user, we compute the share of these outcomes out of their plays and estimate~\eqref{eq:ate_reg} for each. Figure~\ref{fig:match_quality_frac} shows statistically significant and negative effects on the probability of completing at least 30 minutes and on the share of positive thumbs, precise nulls on the probabilities of the 15- and 45-minute completion shares and the positive thumbs ratio, and insignificant effects on the share of negative thumbs.  From our conceptual model, these results are a combination of the counteracting forces of extensive-margin entry and inframarginal switching. To isolate the latter, we temporally order each user's played titles and plot the match quality measures for the $k$-th played title in Appendix Figure~\ref{fig:hazard_match_quality}. Users play titles of lower match quality as they play more titles, suggesting the negative point estimates could be due to playing marginal titles that users would not have played without the improved recommendations. To conservatively account for extensive-margin entry and isolate the effects of inframarginal switching, we compute Lee bounds \citep{lee2009training} on match quality for the $k$-th played title in Figure~\ref{fig:first_k_match_quality} which suggest that we cannot consistently sign whether the newly played titles are higher or lower quality.\footnote{The assumption of monotonicity required for the validity of Lee bounds -- treatment inducing additional consumption compared to the control -- is reasonable given the results in Section~\ref{subsec:overall_consumption}.} Overall, our results indicate that users found more ``high quality" titles to play and that the average title is of comparable quality, especially once we account for extensive-margin selection into additional plays due to the treatment.

\section{Conclusion}

This paper studied how advances in recommendation technology reallocate the concentration of consumption.  Using a unique experiment on Netflix's recommender system encompassing 8.5 million users, we show that such improvements redistribute plays from a small number of superstar titles to a larger number of middle-tail titles while also increasing overall engagement and user reliance on recommendations. These results stand in contrast to previous work that hypothesized that increased personalization would lead to an increasing consumption share of long-tail and superstar titles at the expense of the middle-tail. Our model provides a rationalization for this: in markets with horizontal preference heterogeneity, improving the RecSys allows it to extract more precise signals out of limited data, which attenuates the popularity bias inherent in existing algorithms.

These results have important managerial implications, as they suggest that the returns of middle-tail products to platforms depend on algorithmic maturity and the ability of the RecSys to exploit limited data. As recommendation technology continues to improve, it becomes more feasible for platforms to target an increasingly broad set of middle-tail titles that earlier generations of recommendation technologies were less able to personalize. This frontier remains bounded, however: long-tail titles generate very sparse interaction data, making targeting difficult for even mature technologies. Furthermore, it may be difficult to predict \emph{ex ante} which titles will fall into this margin.  Even so, our findings reframe the long-running debate over whether recommendation systems concentrate or disperse consumption: the answer is neither fixed nor universal but evolves with algorithmic maturity.  At the current frontier, continued improvement lifts overall engagement and reallocates it toward the middle-tail.

\bibliographystyle{chicago}
\bibliography{bib}

\newpage

 \counterwithin{figure}{section} 
\counterwithin{table}{section}
\appendix
\section*{Online Appendix}

\section{Additional Results, Figures and Tables}\label{sec:additional_results}

\begin{figure}[H]
    \caption{Confusion Matrix of Title Bucket Assignment across Treatments}\label{fig:bucket_stability}
    \begin{center}
    \includegraphics[width=0.6\linewidth]{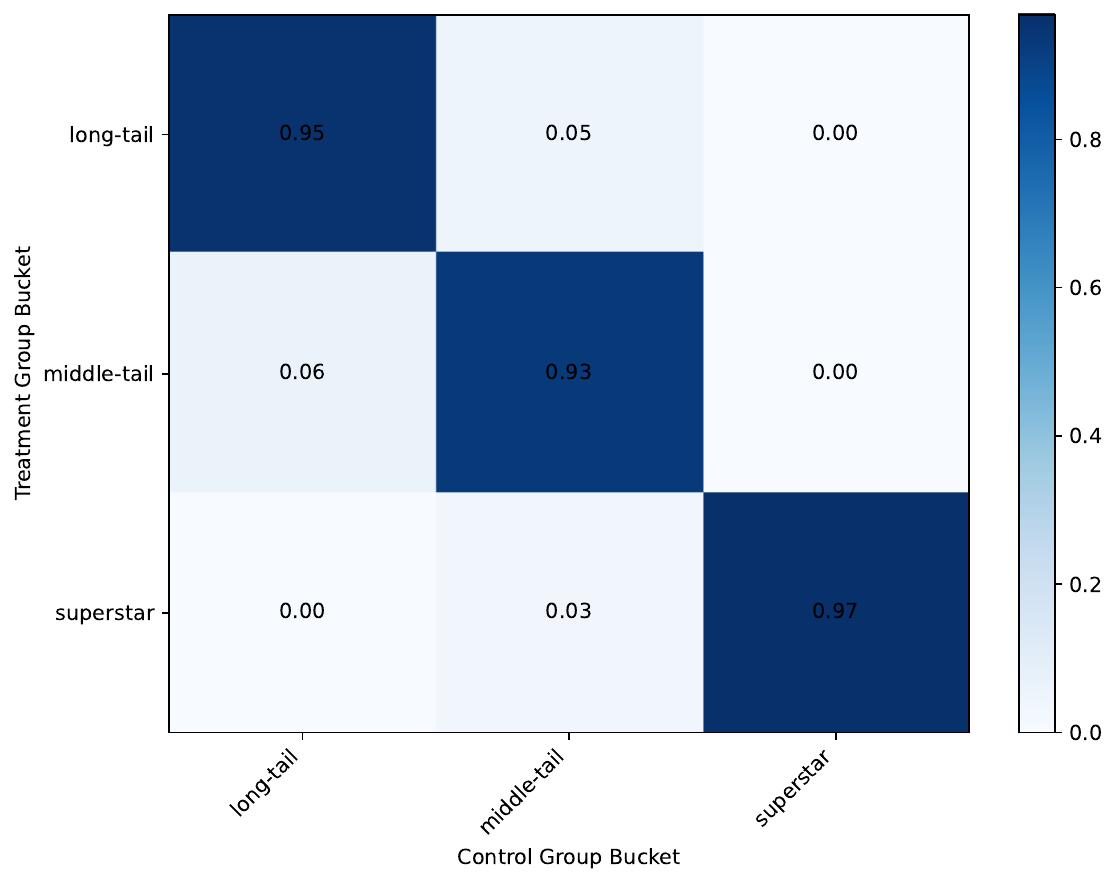}
    \end{center}
    \caption*{\footnotesize \textsc{Notes}: This figure shows the confusion matrix for title categorization into long-tail, middle-tail, and superstars. For each treatment group, we compute each title’s percentile rank in the empirical play distribution within that group. We then classify titles as long-tail if they fall in the bottom 50\% of the distribution, middle-tail if they fall in the next 45\%, and superstars if they fall in the top 5\%. The confusion matrix shows the agreement in categorization across groups.}
\end{figure}

\begin{figure}[H]
    \caption{Recommendation Percentile-Bin Treatment Effects}\label{fig:recommendation_quantile_treatment_effects}
    \begin{center}
    \includegraphics[width=0.6\linewidth]{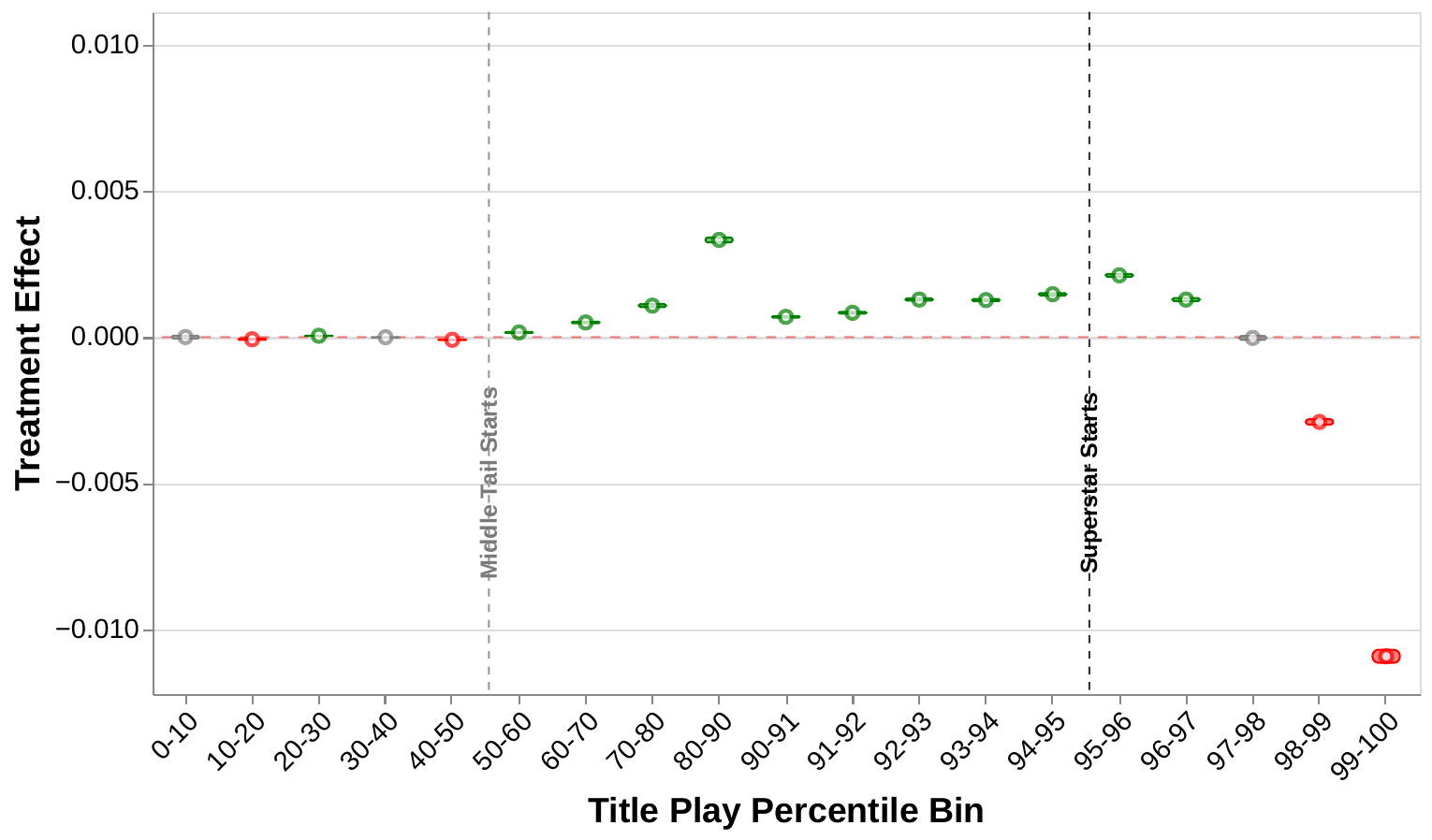}
    \end{center}
    \caption*{\footnotesize \textsc{Notes}: Estimated ATEs on recommendation share within percentile buckets defined using the pooled play distribution across the treatment and control groups.}
\end{figure}

\begin{figure}[ht]
    \centering
    \caption{Treatment Effects on Recommendation Shares (Control Group Ordering)}
    \label{fig:rec-shares-control-only}
    \begin{subfigure}[b]{0.3\linewidth}
        \centering
        \includegraphics[height=2in]{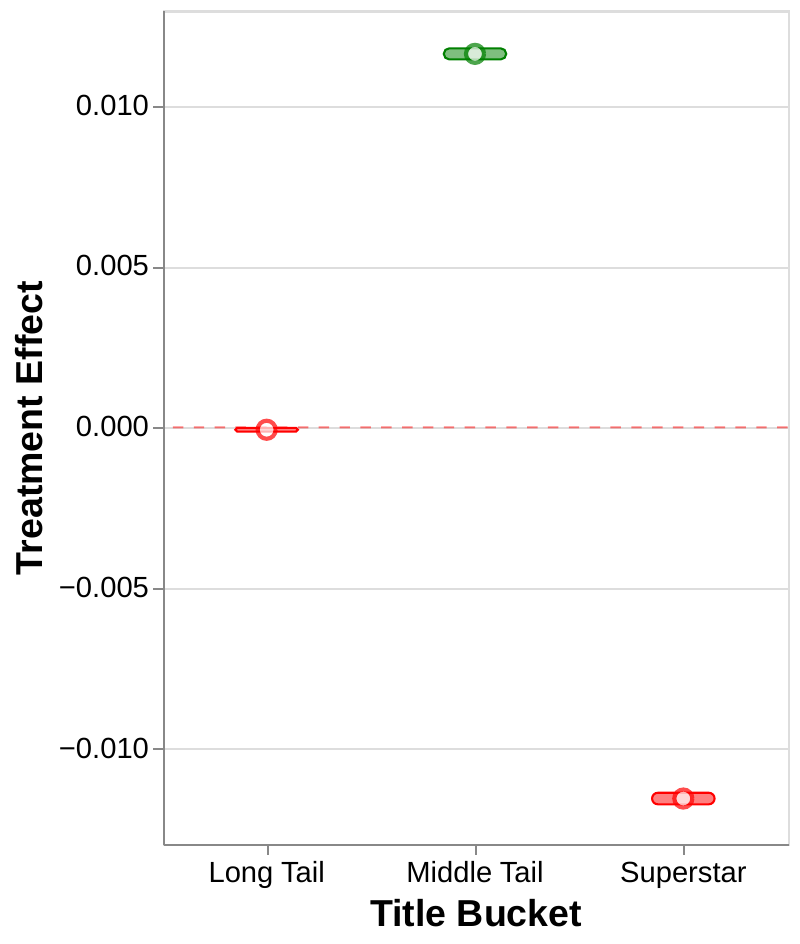}
        \caption{Shares by Bucket}
        \label{fig:rec-shares-avg-control-only}
    \end{subfigure}
    \hfill
    \begin{subfigure}[b]{0.65\linewidth}
        \centering
        \includegraphics[height=2in]{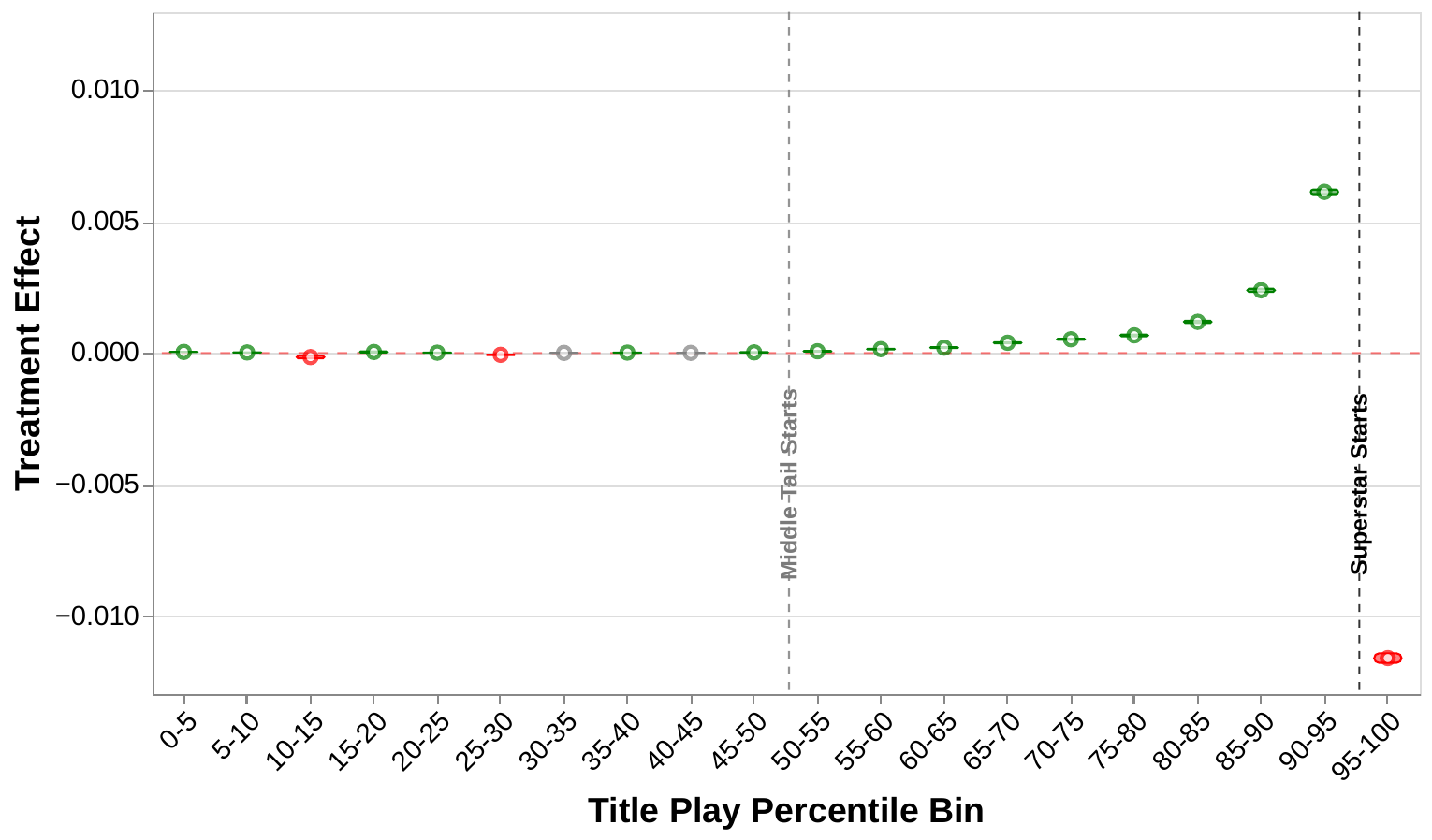}
        \caption{Shares by Ventile Bins}
        \label{fig:rec-shares-ventiles-control-only}
    \end{subfigure}
 
    \caption*{\footnotesize \textsc{Notes}: Estimated ATEs on the distribution of recommendations. Panel (a) presents coefficient estimates from~\eqref{eq:across_groups} with the dependent variable of the share of an individual's recommendations accounted for by each bucket. Panel (b) presents treatment effects on recommendation share within each ventile. In both panels the buckets and ventiles are defined using the control group play distribution. Bars denote 95\% confidence intervals.}
\end{figure}
 
\begin{figure}[H]
    \caption{Play Percentile-Bin Treatment Effects}\label{fig:consumption_quantile_treatment_effects}
    \begin{center}
    \includegraphics[width=0.6\linewidth]{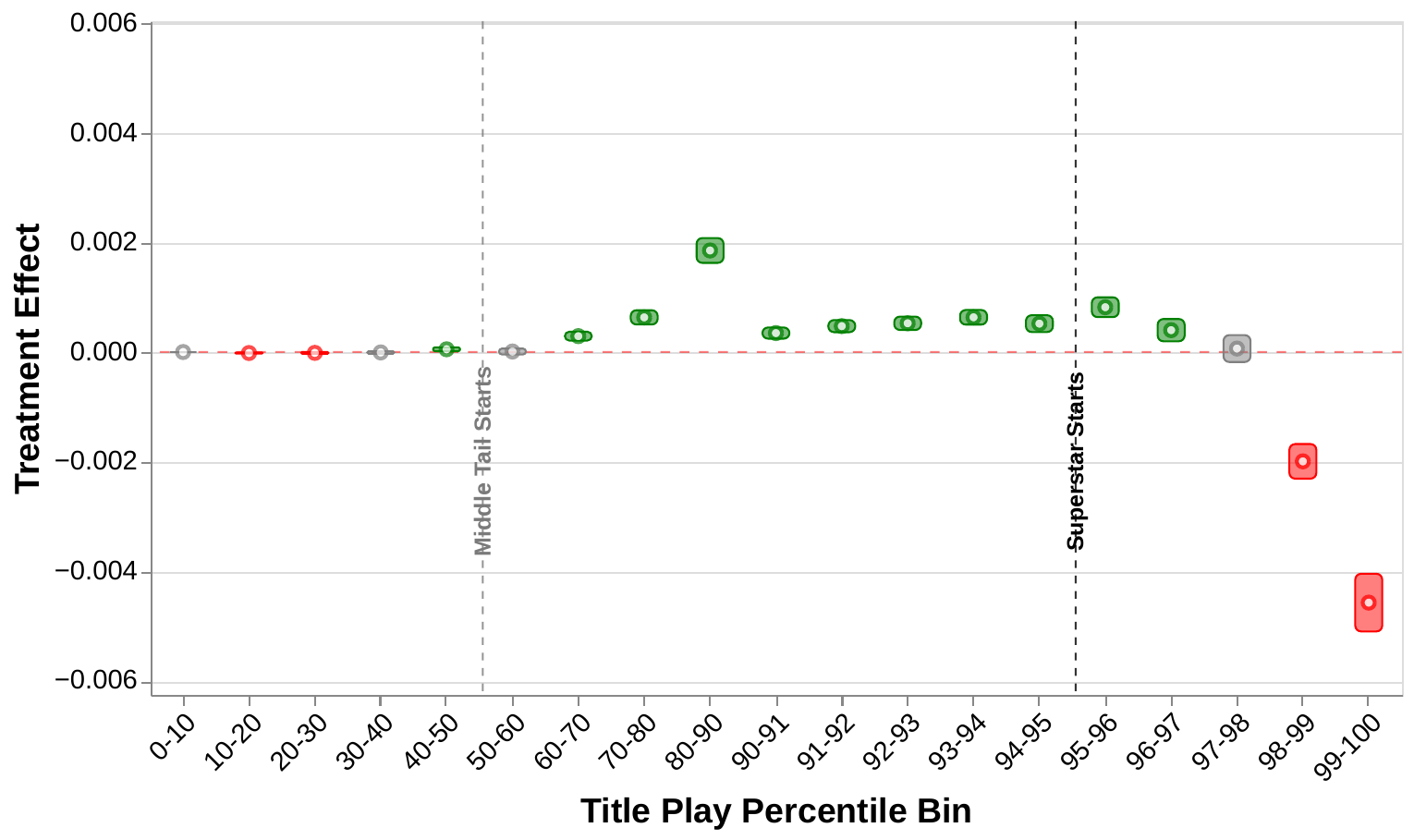}
    \end{center}
    \caption*{\footnotesize \textsc{Notes}: Estimated ATEs on play share within the percentile buckets defined using the pooled play distribution across the treatment and control groups.}
\end{figure}
\begin{figure}[H]
    \centering
    \caption{Treatment Effects on Play Shares (Control Group Ordering)}
    \label{fig:consumption-shares-control-only}
    \begin{subfigure}[t]{0.3\linewidth}
        \centering
        \includegraphics[height=2in]{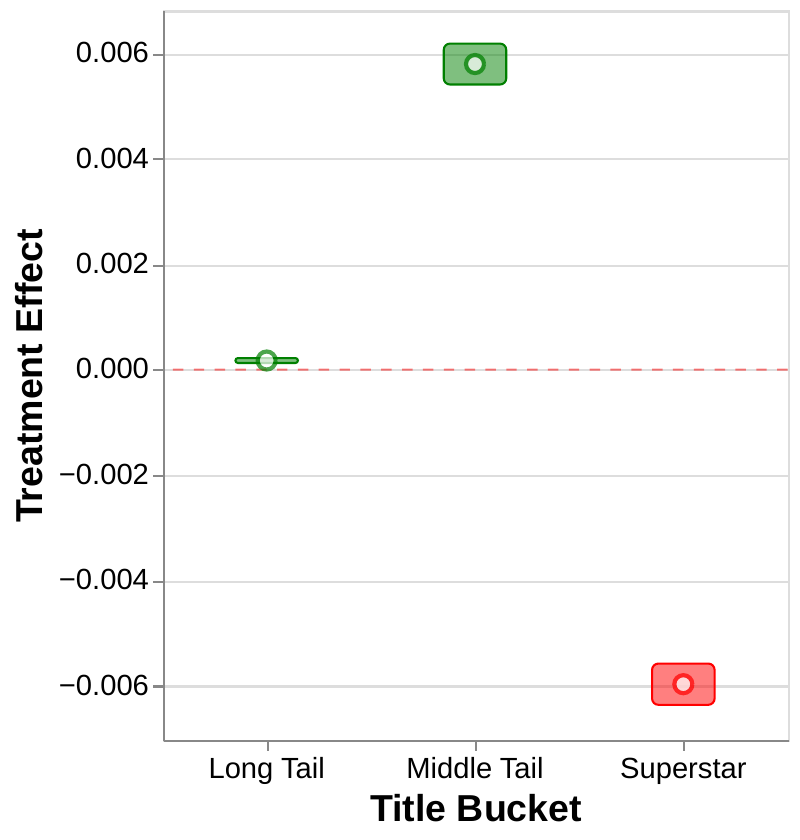}
        \caption{Shares by Bucket}
        \label{fig:consumption-shares-avg-control-only}
    \end{subfigure}
    \hfill
    \begin{subfigure}[t]{0.65\linewidth}
        \centering
        \includegraphics[height=2in]{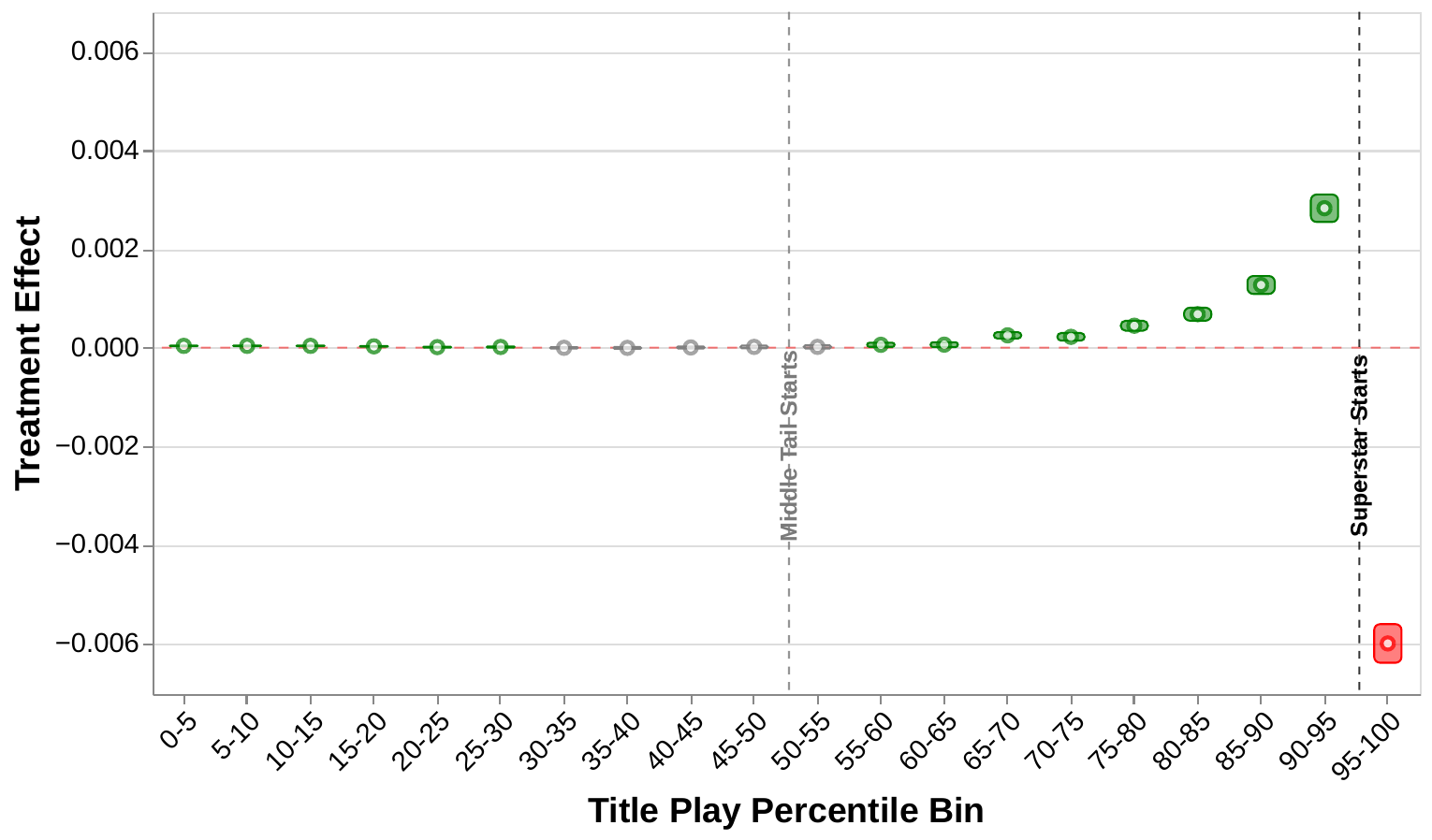}
        \caption{Shares by Ventile Bins}
        \label{fig:consumption-shares-ventiles-control-only}
    \end{subfigure}
 
    \caption*{\footnotesize \textsc{Notes}: Estimated ATEs on the distribution of title plays during the experiment. Panel (a) presents coefficient estimates from specification~\eqref{eq:across_groups} with the dependent variable of the share of an individual's plays accounted for by each bucket. Panel (b) presents treatment effects on play share for each pooled ventile.  In both panels the buckets and ventiles are defined using the control group play distribution. Bars denote 95\% confidence intervals.}
\end{figure}

\begin{figure}[H]
\caption{Overall Engagement Treatment Effects over Time}\label{fig:over_time_treatment_effects}
\begin{center}
\begin{subfigure}[t]{0.32\linewidth}
    \centering
    \includegraphics[width=\linewidth]{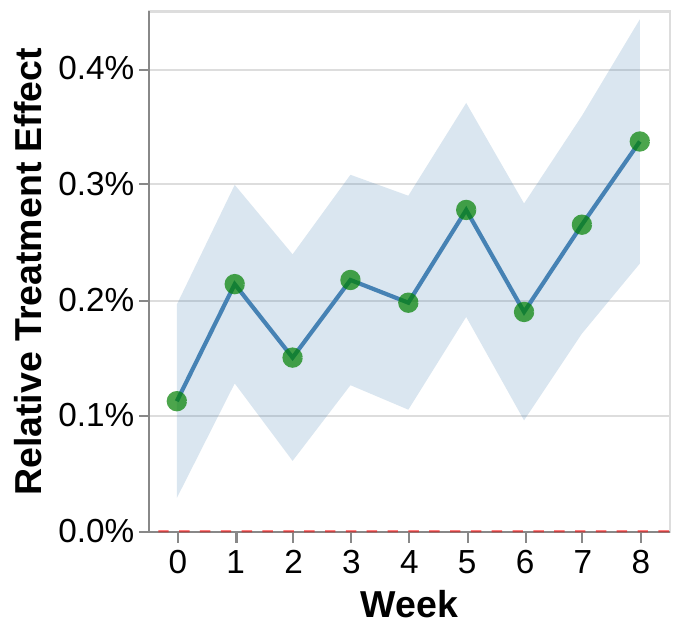}
    \caption{Days with Play}
\end{subfigure}
\begin{subfigure}[t]{0.32\linewidth}
    \centering
    \includegraphics[width=\linewidth]{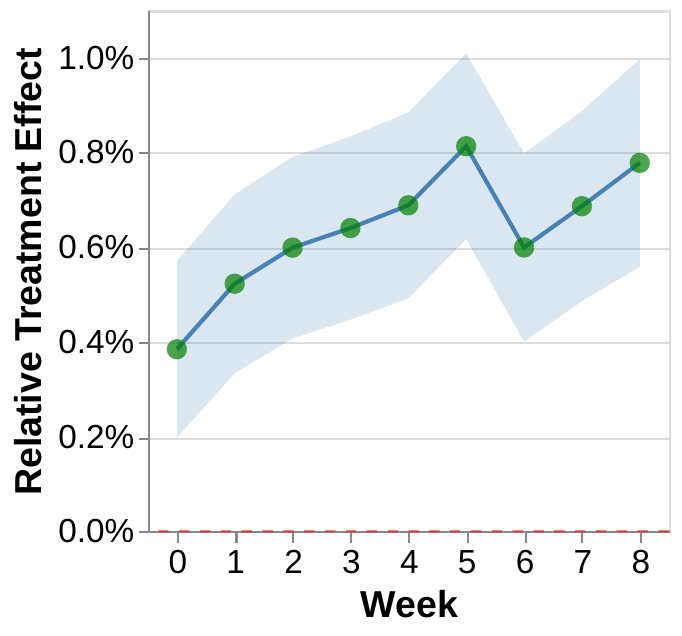}
    \caption{Overall Plays}
\end{subfigure}
\begin{subfigure}[t]{0.32\linewidth}
    \centering
    \includegraphics[width=\linewidth]{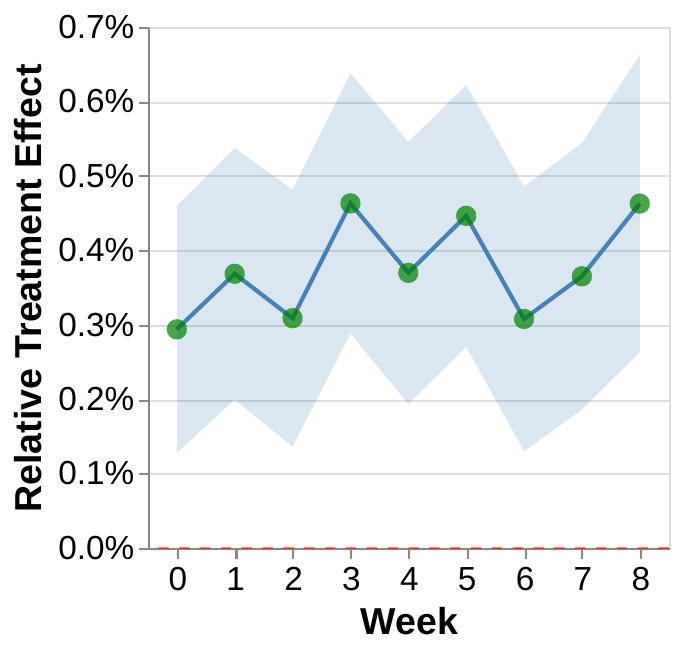}
    \caption{View Hours}
\end{subfigure}
\end{center}
\caption*{\footnotesize \textsc{Notes}: Estimated ATEs from specification~\eqref{eq:ate_reg} for the three primary engagement outcomes -- days with play, overall plays, and view hours -- estimated separately each week of the experimental intervention. Shaded bands denote 95\% confidence intervals computed via robust standard errors.}
\end{figure}

\begin{figure}[H]
\caption{Engagement by Title Type}\label{fig:engagement_title_type}
\begin{center}
\includegraphics[scale=0.4]{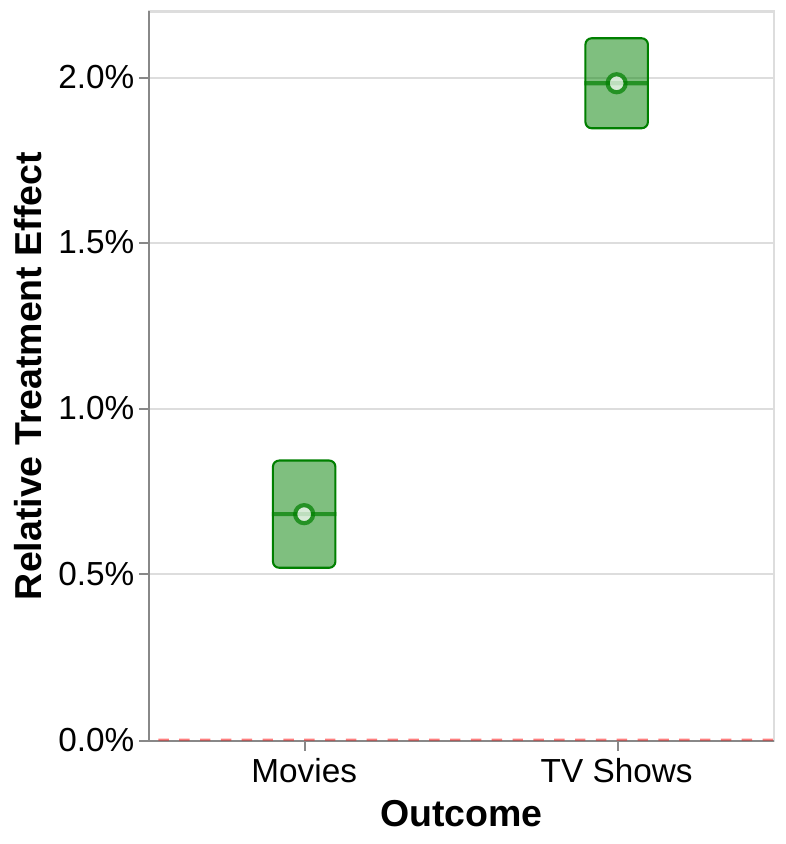}
\end{center}
\caption*{\footnotesize \textsc{Notes}: The figure decomposes the distinct-titles-played in Figure~\ref{fig:overall-engagement} by content type, reporting relative treatment effects on the number of distinct movies and the number of distinct TV shows watched per account. Bars denote 95\% confidence intervals computed via robust standard errors.}
\end{figure}

\begin{figure}[H]
    \caption{Match Quality for First-k Plays}\label{fig:hazard_match_quality}
    \begin{center}
    \begin{subfigure}{0.4\linewidth}
        \centering
        \includegraphics[width=\linewidth]{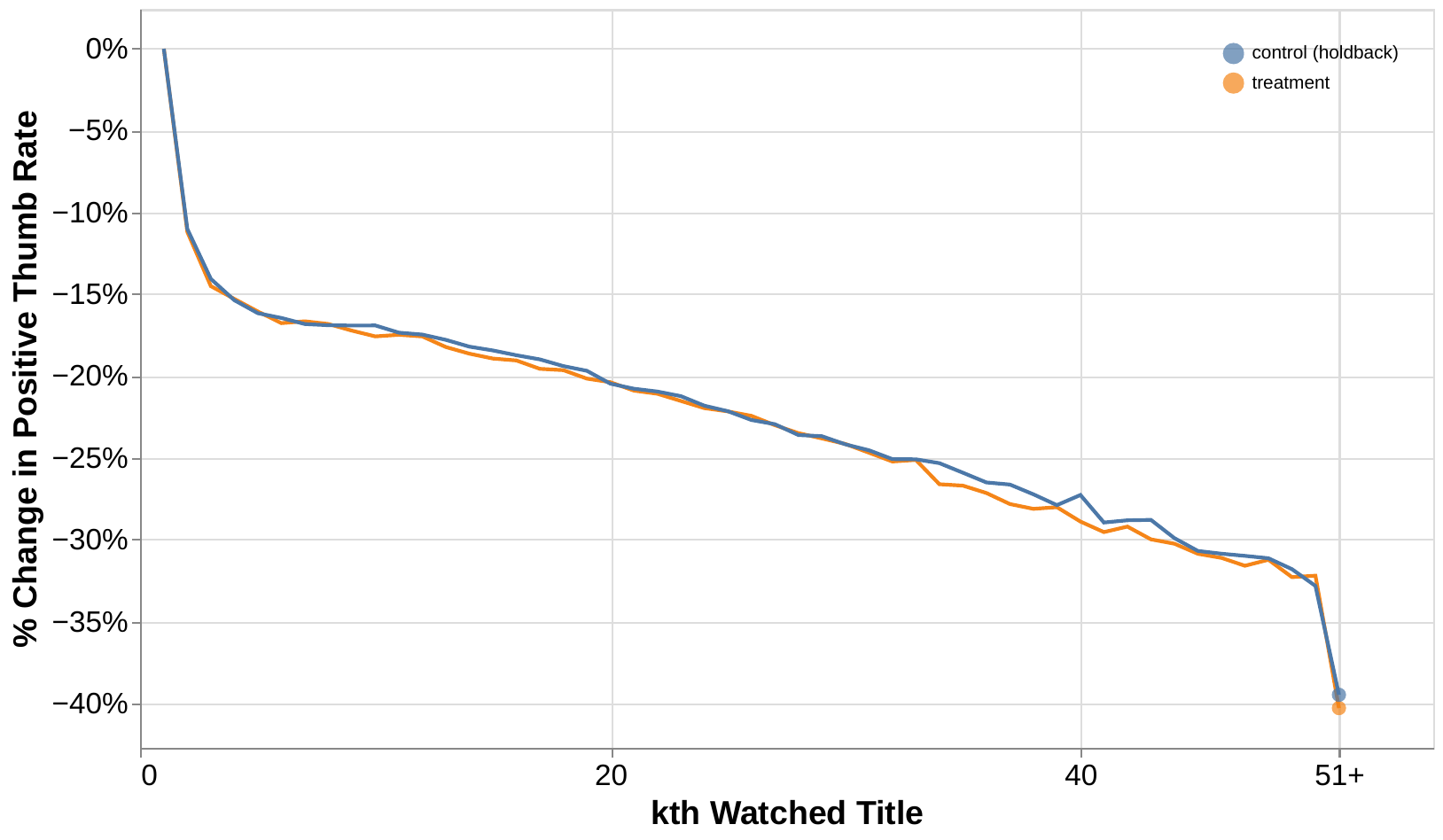}
        \caption{Positive Thumb Rate}\label{fig:hazard_match_quality_thumb_rate}
    \end{subfigure}
    \hfill
    \begin{subfigure}{0.4\linewidth}
        \centering
        \includegraphics[width=\linewidth]{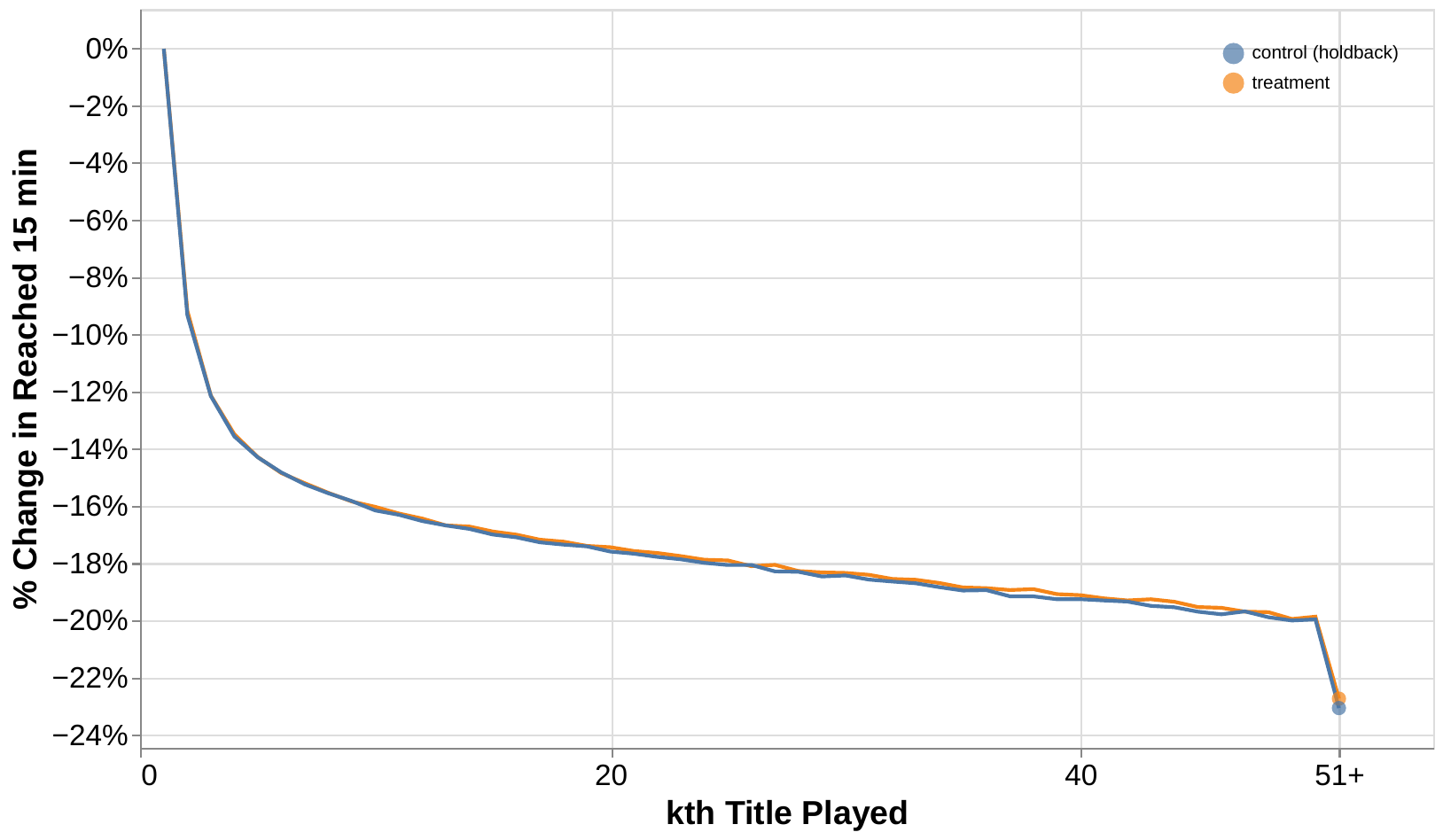}
        \caption{Share Watched At Least 15 Minutes}\label{fig:hazard_match_quality_clear_15m}
    \end{subfigure}

    \vspace{1em}

    \begin{subfigure}{0.4\linewidth}
        \centering
        \includegraphics[width=\linewidth]{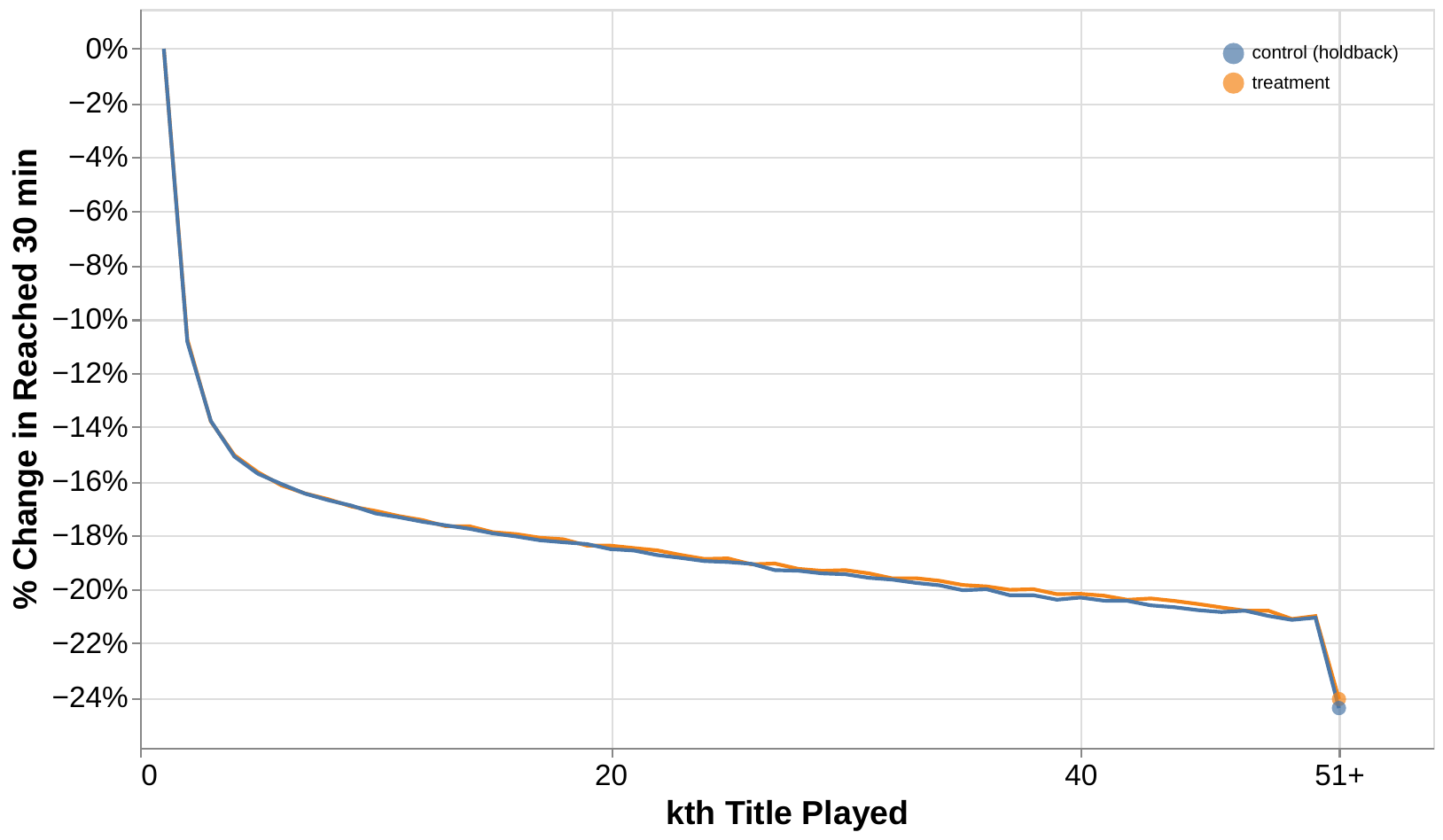}
        \caption{Share Watched At Least 30 Minutes}\label{fig:hazard_match_quality_clear_30m}
    \end{subfigure}
    \hfill
    \begin{subfigure}{0.4\linewidth}
        \centering
        \includegraphics[width=\linewidth]{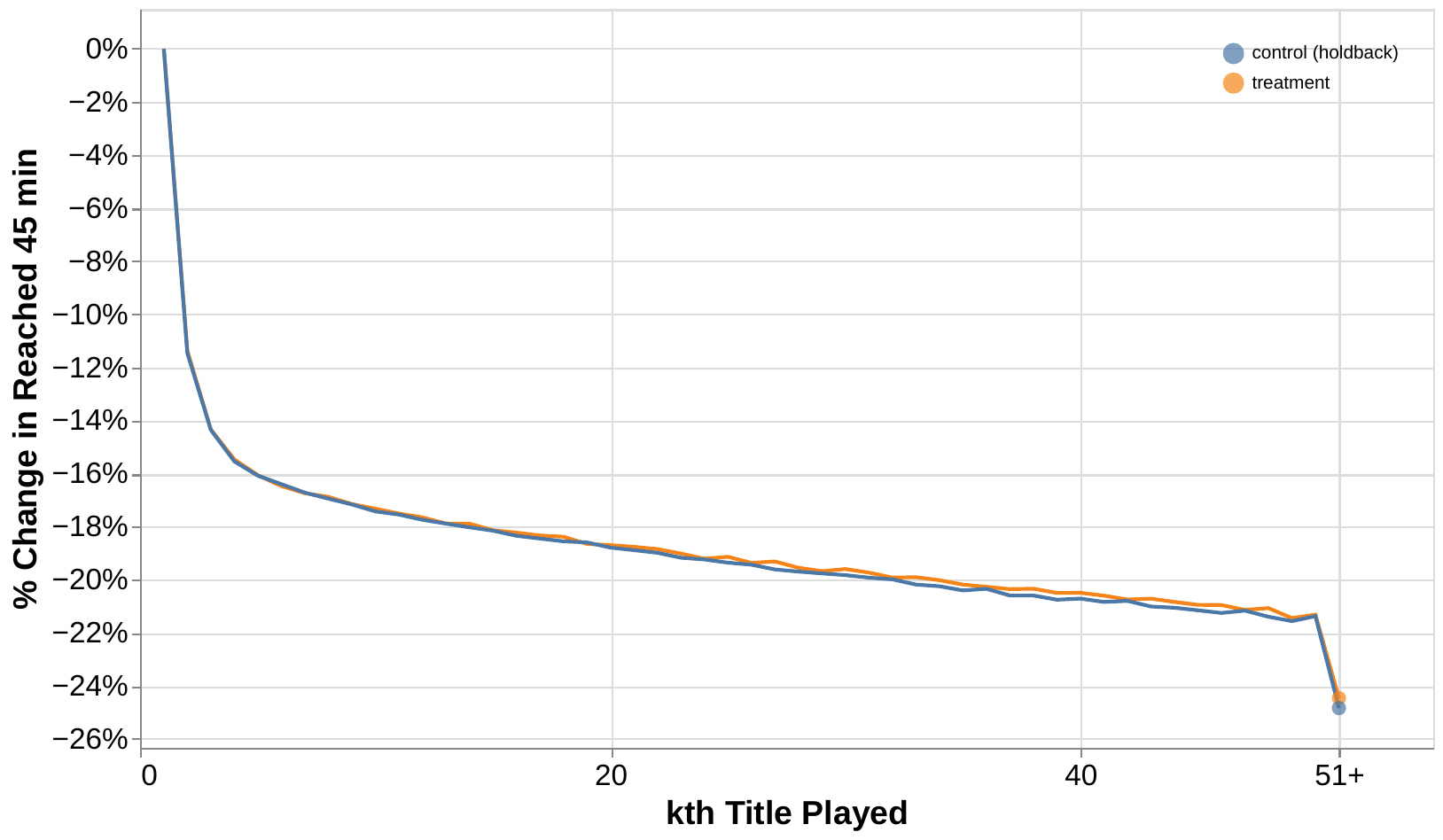}
        \caption{Share Watched At Least 45 Minutes}\label{fig:hazard_match_quality_clear_45m}
    \end{subfigure}
    \end{center}
        \caption*{\footnotesize \textsc{Notes}: The figure plots the underlying level of positive thumb rate (Panel a) and the share of plays watched at least 15, 30, and 45 minutes during the first day of viewing (Panels b--d) against the exact $k$th title played, one line per treatment arm, shown up to $k=50$ and then a single pooled $51$+ point. The positive thumb rate is only computed for titles that were watched sufficiently long to have the thumbs prompt appear. TV shows are considered one title, regardless of how many episodes were watched. For each measure, we report percent deviations from the values for $k = 1$.}
\end{figure}
\begin{figure}[H]
    \caption{Lee Bounds on Match Quality for First-k Plays}\label{fig:first_k_match_quality}
    \begin{center}
    \begin{subfigure}{0.48\linewidth}
        \centering
        \includegraphics[width=\linewidth]{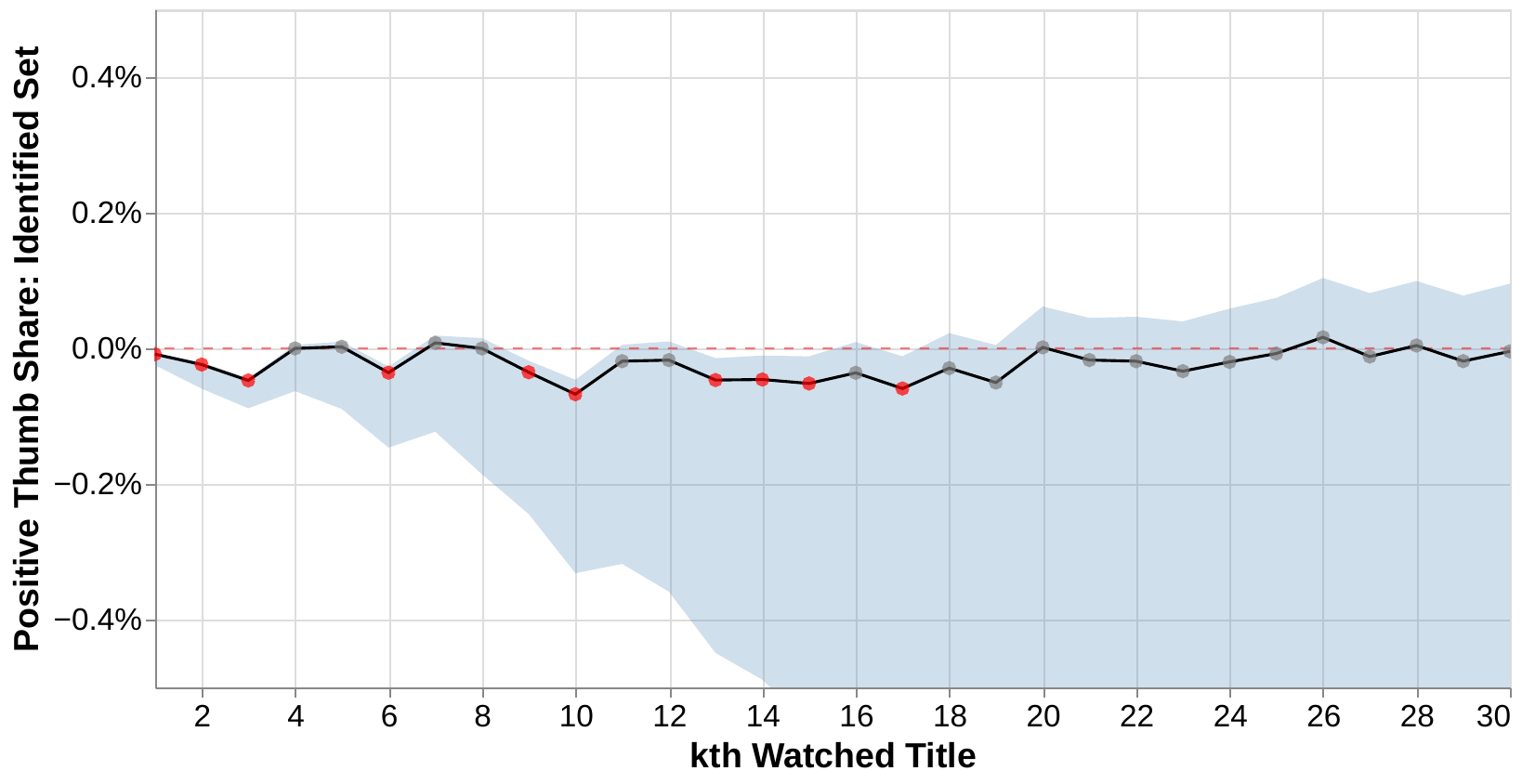}
        \caption{Positive Thumb Share}\label{fig:first_k_match_quality_thumb_share}
    \end{subfigure}
    \hfill
    \begin{subfigure}{0.48\linewidth}
        \centering
        \includegraphics[width=\linewidth]{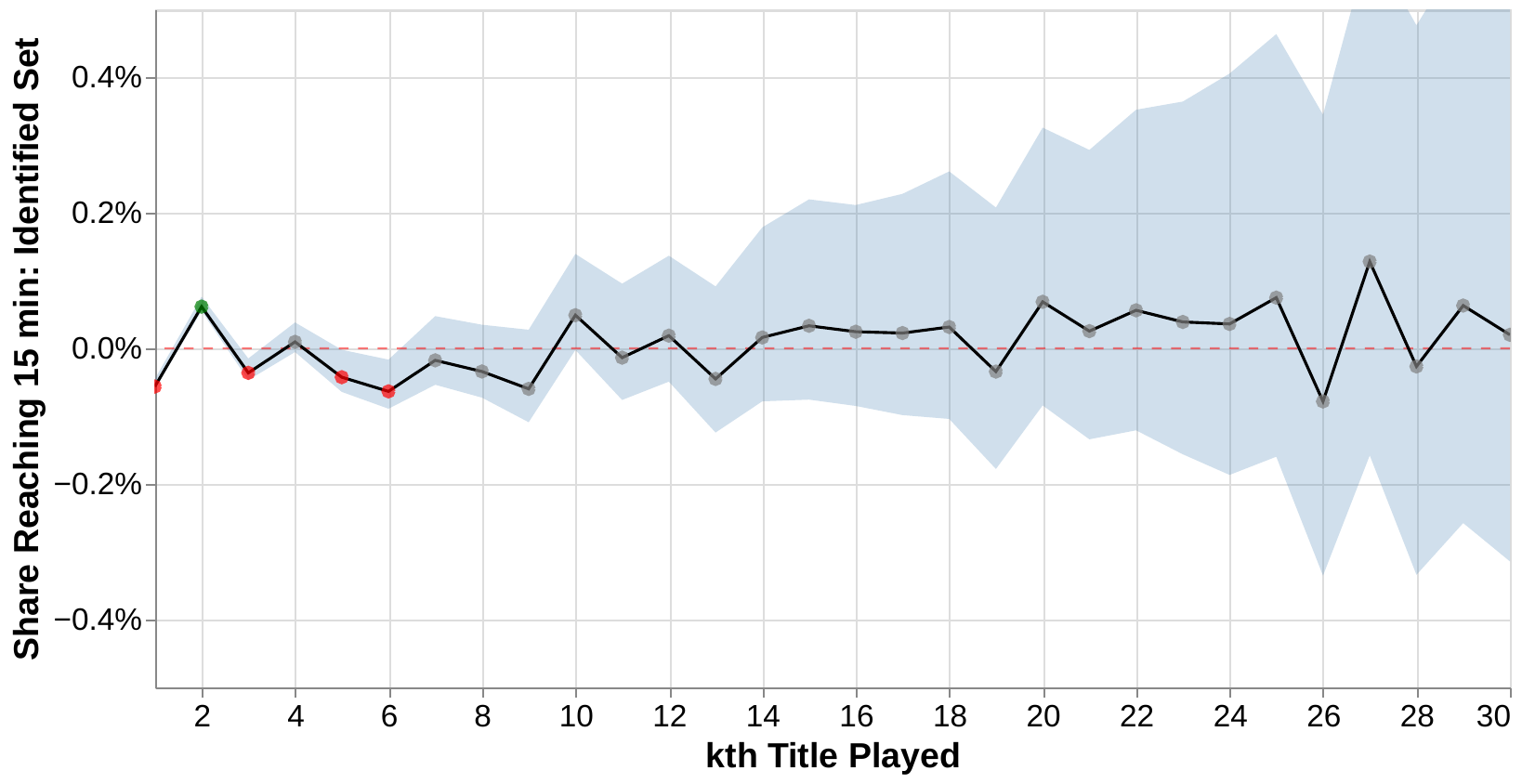}
        \caption{Share Watched At Least 15 Minutes}\label{fig:first_k_match_quality_clear_15m}
    \end{subfigure}

    \vspace{1em}

    \begin{subfigure}{0.48\linewidth}
        \centering
        \includegraphics[width=\linewidth]{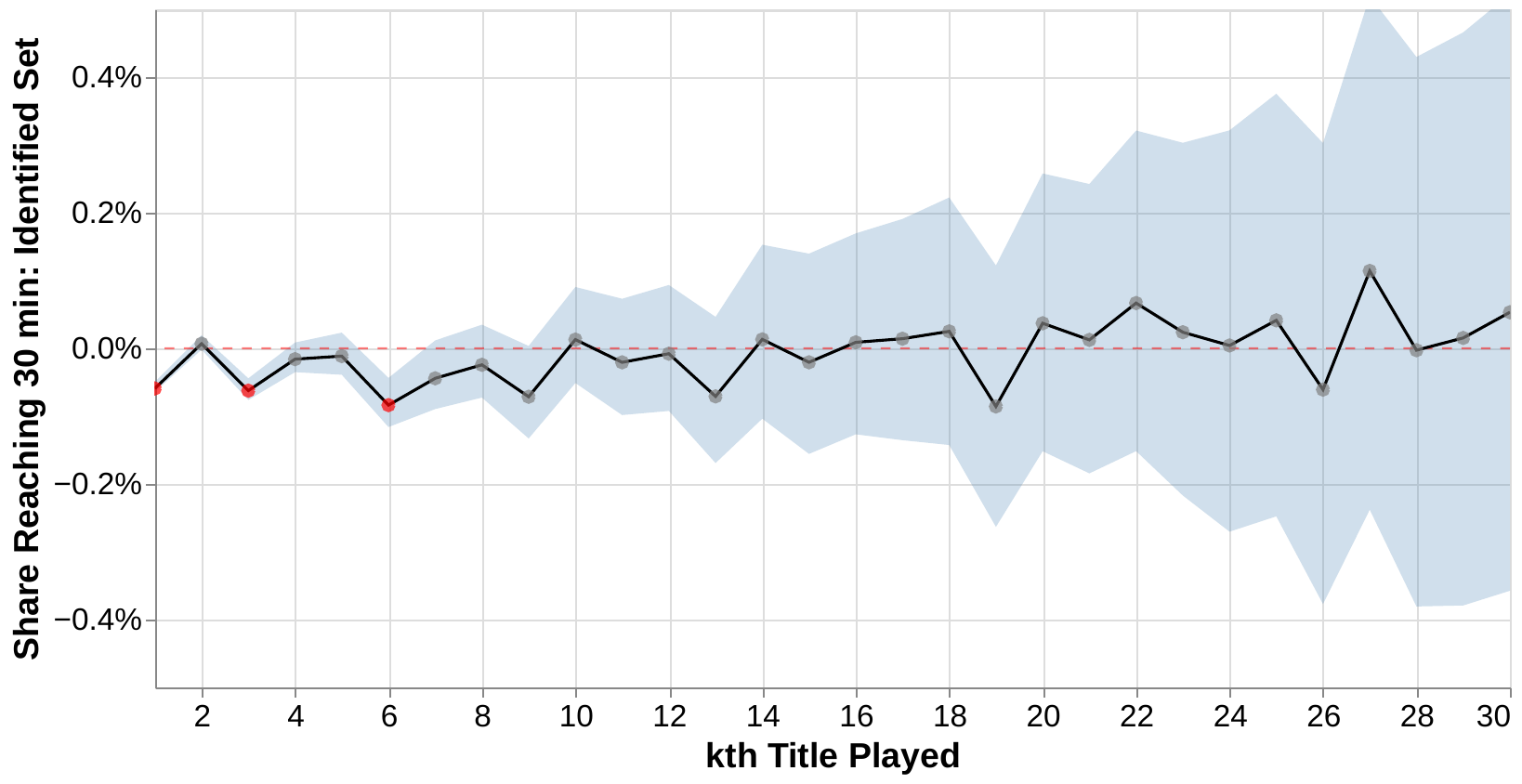}
        \caption{Share Watched At Least 30 Minutes}\label{fig:first_k_match_quality_clear_30m}
    \end{subfigure}
    \hfill
    \begin{subfigure}{0.48\linewidth}
        \centering
        \includegraphics[width=\linewidth]{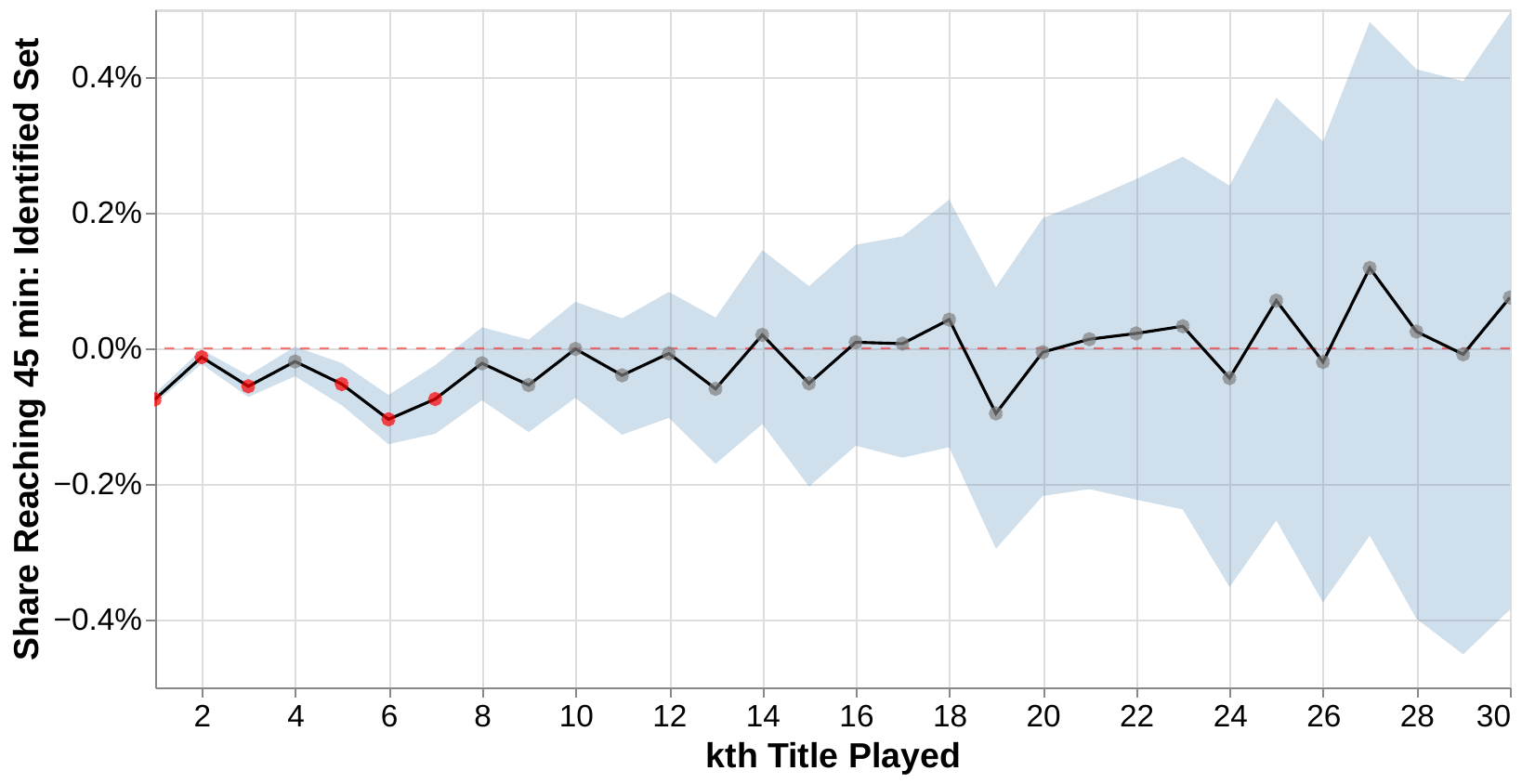}
        \caption{Share Watched At Least 45 Minutes}\label{fig:first_k_match_quality_clear_45m}
    \end{subfigure}
    \end{center}
      \caption*{\footnotesize \textsc{Notes}: The figure presents the estimated Lee bounds \citep{lee2009training} for each $k$-th play on positive thumb share (Panel a) and the share of plays watched at least 15, 30, and 45 minutes during the first day of viewing (Panels b--d). For panel a, $k$ indexes the account's $k$-th distinct title across all content that was watched sufficiently long to trigger the prompt for querying whether a user dislikes the title (negative thumb) or likes the title (single or double positive thumbs up). For panels b--d, $k$ indexes any title that was started. TV shows are considered one title, regardless of how many episodes were watched. The shaded area represents the Lee bounds for each $k$.}
\end{figure}

\section{Omitted Proofs}\label{sec:proofs}

Note that for all the proofs, because the utility shocks have
a continuous density, ties occur with probability zero.

\begin{proof}[Proof of Prediction~\ref{prop:hump}.]
For any parameter $\delta$, and any title $k \in \{L, M, S\}$, the law of total probability applied to the partition $\{j_i^\star = j\}_{j \in \{L, M, S\}}$ gives
\begin{align}\label{equation:proof1abcd}
    f_k(\delta) = \Pr(k \in \mathcal{R}_i(\delta)) = \sum_{j \in \{L, M, S\}}\Pr(k \in \mathcal{R}_i(\delta) \mid j_i^\star = j) \, p_j.
\end{align}
By Assumption~\ref{a:tech} (popularity default), conditional on $j_i^\star = j$ the recommender outputs either $j$ (with probability $q_j(\delta)$) or the superstar $S$ (with probability $1 - q_j(\delta)$).

Writing \eqref{equation:proof1abcd} for $k \in \{L, M, S\}$, we obtain
\begin{align}
    f_L(\delta) & \label{equation:123231234} = p_L \, q_L(\delta), \\
    f_M(\delta) & \label{equation:1232312345} = p_M \, q_M(\delta), \\
    f_S(\delta) & \label{equation:12323123456} =  p_S + p_M \, (1 - q_M(\delta)) + p_L \, (1 - q_L(\delta)).
\end{align}
\eqref{equation:123231234} holds since $L$ is recommended only when $j_i^\star = L$, and similarly for \eqref{equation:1232312345}. \eqref{equation:12323123456} reflects the fact that $S$ is recommended both when $j_i^\star = S$ (in which case $\Pr(S \in \mathcal{R}_i(\delta) \mid j_i^\star = S) = 1$) and when $j_i^\star \in \{L, M\}$ in the case where the recommender defaults to $S$ (which happens respectively with probability $(1 - q_L(\delta))$ and $(1 - q_M(\delta))$).

By Assumption~\ref{a:tech} (monotonicity), we can differentiate $f_L, f_M$, and $f_S$ with respect to $\delta$, which gives that
\begin{align*}
    f_M'(\delta) - f_L'(\delta) &= p_M \, q_M'(\delta) - p_L \, q_L'(\delta) > 0, \\
    f_S'(\delta) - f_L'(\delta) &= -p_M \, q_M'(\delta) - p_L \, q_L'(\delta) - p_L q_L'(\delta) < 0,
\end{align*}
where both inequalities follow from Assumption~\ref{a:tech}, hence the result.
\end{proof}
\begin{proof}[Proof of Prediction~\ref{prop:shares}.]
We defined $\pi_k(\delta) = \Pr(c_i = k)$ as the probability that user $i$ consumes title $k$. Applying the law of total probability to the partition $\{j_i^\star = j\}_{j \in \{L, M, S\}}$, we can write
\begin{align*}
    \pi_k(\delta) = \sum_{j \in \{L, M, S\}} p_j \, \Pr(c_i = k \mid j_i^\star = j) .
\end{align*}
For $j \in \{L, M\}$, conditional on $j_i^\star = j$ the recommender outputs $j$ with probability $q_j(\delta)$ and $S$ with probability $1 - q_j(\delta)$; for $j = S$, the recommender always outputs $S$ regardless of $\delta$. Therefore, we have that for $k \in \{L, M, S\}$
\begin{align*}
    \Pr(c_i = k \mid j_i^\star = j) = q_j(\delta) \, \Pr(c_i = k \mid j_i^\star = j, \mathcal{R}_i = \{j\}) + (1 - q_j(\delta)) \Pr(c_i = k \mid j_i^\star = j, \mathcal{R}_i = \{S\}) .
\end{align*}
The conditional consumption probabilities $\Pr(c_i = k \mid j_i^\star = j, \mathcal{R}_i = \{r\})$ depend only on $(U_{iL}, U_{iM}, U_{iS}, b)$, not on $\delta$. Differentiating, we obtain
\begin{align}\label{eq:pi_deriv}
    \pi_k'(\delta) = \sum_{j \in \{L, M\}} p_j \, q_j'(\delta) \big[ \Pr(c_i = k \mid j_i^\star = j, \mathcal{R}_i = \{j\}) - \Pr(c_i = k \mid j_i^\star = j, \mathcal{R}_i = \{S\}) \big].
\end{align}
We use a key observation to simplify \eqref{eq:pi_deriv}. Conditional on $j_i^\star = j$, we have $U_{ij} \geq U_{ik}$ for all $k$. Under $\mathcal{R}_i = \{j\}$, the bonus on $j$ reinforces this ranking, so the user consumes either $j$ or the outside option. Under $\mathcal{R}_i = \{S\}$, the bonus on $S$ can lift $S$ above $j$ but cannot lift any third title above $j$, so the user consumes either $j$, $S$, or the outside option. In particular, for $k \notin \{j, S\}$,
\begin{align*}
    \Pr(c_i = k \mid j_i^\star = j, \mathcal{R}_i = \{j\}) = \Pr(c_i = k \mid j_i^\star = j, \mathcal{R}_i = \{S\}) = 0.
\end{align*}
The bracketed term in \eqref{eq:pi_deriv} is therefore nonzero only when $k \in \{j, S\}$.

This observation partitions the analysis into two structurally distinct cases. When $k$ is a non-default title ($k \in \{L, M\}$), only the $j = k$ term in \eqref{eq:pi_deriv} survives: the only way a precision gain changes the consumption of $k$ is by correctly identifying a $k$-type user and redirecting them from the popularity default $S$ to their best title $k$. When $k$ is the popularity default ($k = S$), both $j = L$ and $j = M$ terms contribute: every precision gain redirects mass away from $S$ as some $L$- and $M$-type users are correctly identified and shifted away. We handle these two cases in turn.

\textbf{Case $k \in \{L, M\}$.} Only the $j = k$ term contributes:
\begin{align*}
    \pi_k'(\delta) = p_k \, q_k'(\delta) \big[ \Pr(c_i = k \mid j_i^\star = k, \mathcal{R}_i = \{k\}) - \Pr(c_i = k \mid j_i^\star = k, \mathcal{R}_i = \{S\}) \big].
\end{align*}
Conditional on $\{j_i^\star = k\}$, the bonus on $k$ under $\mathcal{R}_i = \{k\}$ makes $c_i = k$ weakly more likely than under $\mathcal{R}_i = \{S\}$, where the bonus instead pushes consumption toward $S$. To establish strict inequality, consider the event $\{j_i^\star = k,\, -b < U_{ik} < 0,\, U_{iS} + b < 0\}$, which has positive measure by the bounded, strictly positive density of $\theta$. On this event:
\begin{itemize}
    \item Under $\mathcal{R}_i = \{S\}$, $U_{iS} + b < 0$ and $U_{ik} < 0$, so the user takes the outside option and $c_i = 0$.
    \item Under $\mathcal{R}_i = \{k\}$, $V_{ik} = U_{ik} + b > 0$, so the user consumes $k$.
\end{itemize}
Hence the bracketed term is strictly positive. By Assumption~\ref{a:tech}, $p_M q_M'(\delta) > 0$ and $p_L q_L'(\delta) \geq 0$, hence $\pi_M'(\delta) > 0$ and $\pi_L'(\delta) \geq 0$.

\textbf{Case $k = S$.} Both $j = L$ and $j = M$ contribute:
\begin{align*}
    \pi_S'(\delta) = \sum_{j \in \{L, M\}} p_j \, q_j'(\delta) \big[ \Pr(c_i = S \mid j_i^\star = j, \mathcal{R}_i = \{j\}) - \Pr(c_i = S \mid j_i^\star = j, \mathcal{R}_i = \{S\}) \big].
\end{align*}
Conditional on $j_i^\star = j$ for $j \in \{L, M\}$, the bonus under $\mathcal{R}_i = \{j\}$ is on $j$ rather than $S$, so $S$ is weakly less likely to be consumed than under $\mathcal{R}_i = \{S\}$. Each bracketed term is nonpositive. 

The $j = M$ term is strictly negative. Consider the event $\{j_i^\star = M,\, 0 < U_{iM} - U_{iS} < b,\, U_{iS} + b > 0\}$, which has positive measure by the bounded, strictly positive density of $\theta$. On this event:
\begin{itemize}
    \item Under $\mathcal{R}_i = \{S\}$, the bonus lifts $U_{iS} + b$ above $U_{iM}$ (since $U_{iM} - U_{iS} < b$) and above $U_{iL}$ (since $U_{iL} \leq U_{iM}$) and above $0$ (since $U_{iS} + b > 0$), so the user consumes $S$.
    \item Under $\mathcal{R}_i = \{M\}$, the bonus is on $M$ instead, so $V_{iM} = U_{iM} + b > V_{iS} = U_{iS}$, and the user does not consume $S$.
\end{itemize}
Hence $\Pr(c_i = S \mid j_i^\star = M, \mathcal{R}_i = \{S\}) > \Pr(c_i = S \mid j_i^\star = M, \mathcal{R}_i = \{M\})$, and the $j = M$ bracketed term in $\pi_S'(\delta)$ is strictly negative. Since $p_M q_M'(\delta) > 0$ by Assumption~\ref{a:tech}, we obtain $\pi_S'(\delta) < 0$.
\end{proof}

\begin{proof}[Proof of Prediction~\ref{prop:engagement}.]
The overall strategy is to decompose engagement by the user's best title $j_i^\star$ and show that improvements in $\delta$ raise the engagement of $L$- and $M$-type users by redirecting them from the popularity default $S$ to their true best title, where the recommendation bonus $b > 0$ makes consumption more likely.

Let $E(\delta) \coloneqq \Pr(c_i \neq 0)$ denote the engagement probability. Applying the law of total probability to the partition $\{j_i^\star = j\}_{j \in \{L, M, S\}}$, we can write
\begin{align*}
    E(\delta) = \Pr(c_i \neq 0) = \sum_{j \in \{L, M, S\}} p_j \, \Pr(c_i \neq 0 \mid j_i^\star = j) .
\end{align*}
The $S$-type term contributes nothing to $E'(\delta)$: when $j_i^\star = S$, the popularity default coincides with the user's best title, hence $\mathcal{R}_i = \{S\}$ with probability one regardless of $\delta$. Therefore $\Pr(c_i \neq 0 \mid j_i^\star = S) = \Pr(U_{iS}+b>0\mid j_i^\star=S)$ is independent of $\delta$.

The main consequence of the increase in $\delta$ comes from the $L$- and $M$-type terms, where $\delta$ shifts the recommendation away from the default $S$ toward the user's actual best title. 

For $j \in \{L, M\}$, conditional on $j_i^\star = j$ the recommender outputs $j$ with probability $q_j(\delta)$ and $S$ with probability $1 - q_j(\delta)$:
\begin{align*}
    \Pr(c_i \neq 0 \mid j_i^\star = j) = q_j(\delta) \, \Pr(c_i \neq 0 \mid j_i^\star = j, \mathcal{R}_i = \{j\}) + (1 - q_j(\delta)) \, \Pr(c_i \neq 0 \mid j_i^\star = j, \mathcal{R}_i = \{S\}).
\end{align*}
The conditional consumption probabilities given $\mathcal{R}_i$ are functions of $(U_{iL}, U_{iM}, U_{iS}, b)$ alone and do not depend on $\delta$. Differentiating with respect to $\delta$ gives that
\begin{align*}
    \frac{d}{d\delta} \Pr(c_i \neq 0 \mid j_i^\star = j) = q_j'(\delta) \big[ \Pr(c_i \neq 0 \mid j_i^\star = j, \mathcal{R}_i = \{j\}) - \Pr(c_i \neq 0 \mid j_i^\star = j, \mathcal{R}_i = \{S\}) \big].
\end{align*}
Since the $S$-type term in $E(\delta)$ is constant in $\delta$, multiplying by $p_j$ and summing over $j \in \{L, M\}$ gives that
\begin{align}\label{eq:E_deriv}
    E'(\delta) = \sum_{j \in \{L, M\}} p_j \, q_j'(\delta) \big[ \Pr(c_i \neq 0 \mid j_i^\star = j, \mathcal{R}_i = \{j\}) - \Pr(c_i \neq 0 \mid j_i^\star = j, \mathcal{R}_i = \{S\}) \big].
\end{align}
Each term measures the engagement gain from redirecting a $j$-type user's recommendation away from $S$ toward $j$, weighted by the marginal precision gain $p_j q_j'(\delta)$.

We now show that each redirection gain is strictly positive. For any $j \in \{L, M\}$, conditional on $\{j_i^\star = j\}$, we have $U_{ij} \geq U_{iS}$, thus recommending $j$ (which gets the bonus) gives a higher decision utility $V_{ij}$ than recommending $S$. The recommendation bonus thus makes consumption weakly more likely under $\mathcal{R}_i = \{j\}$. To get strict inequality, we identify a set of users for whom the recommendation directly flips the consumption decision: those whose true best title $j$ has utility just below the outside option, while $S$ remains below even with the bonus. Since the $\theta_{ij}$ have bounded, strictly positive density on $\mathbb{R}$,
\begin{align*}
    \Pr\!\big( -b < U_{ij} \leq 0,\; U_{iS} + b \leq 0 \,\big|\, j_i^\star = j \big) > 0.
\end{align*}
On this event, $U_{ik} \leq U_{ij} \leq 0$ for all $k$ and $U_{iS} + b \leq 0$, hence under $\mathcal{R}_i = \{S\}$ the user takes the outside option ($c_i = 0$), while under $\mathcal{R}_i = \{j\}$ the bonus pushes $V_{ij} = U_{ij} + b > 0$ and the user consumes $j$ ($c_i \neq 0$). Therefore
\begin{align*}
    \Pr(c_i \neq 0 \mid j_i^\star = j, \mathcal{R}_i = \{j\}) > \Pr(c_i \neq 0 \mid j_i^\star = j, \mathcal{R}_i = \{S\}).
\end{align*}
By Assumption~\ref{a:tech}, $p_L q_L'(\delta) \geq 0$ and $p_M q_M'(\delta) > 0$. Combining it with the bracketed term in \eqref{eq:E_deriv} being strictly positive, the $L$-type term is nonnegative and the $M$-type term is strictly positive.
Therefore $E'(\delta) > 0$.
\end{proof}

\begin{proof}[Proof of Prediction~\ref{prop:recdep}.]
The strategy is to show that improvements in $\delta$ raise the probability of recommendation-driven consumption faster than they raise overall engagement, so the conditional share of consumption coming from recommendations rises.

Let $\rho(\delta) \coloneqq \Pr(c_i \in \mathcal{R}_i \mid c_i \neq 0)$ denote the recommendation share. Since $\{c_i \in \mathcal{R}_i\} \subset \{c_i \neq 0\}$,
\begin{align*}
    \rho(\delta) = \frac{\Pr(c_i \in \mathcal{R}_i, c_i \neq 0)}{\Pr(c_i \neq 0)} = \frac{\Pr(c_i \in \mathcal{R}_i)}{E(\delta)}.
\end{align*}
We first show that $\frac{d}{d\delta} \Pr(c_i \in \mathcal{R}_i) \geq E'(\delta)$. From \eqref{eq:E_deriv} in the proof of Prediction~\ref{prop:engagement}, we can write
\begin{align}\label{eq:rho_den}
    E'(\delta) = \sum_{j \in \{L, M\}} p_j \, q_j'(\delta) \big[ \Pr(c_i \neq 0 \mid j_i^\star = j, \mathcal{R}_i = \{j\}) - \Pr(c_i \neq 0 \mid j_i^\star = j, \mathcal{R}_i = \{S\}) \big].
\end{align}
An analogous decomposition of $\Pr(c_i \in \mathcal{R}_i)$ gives
\begin{align}\label{eq:rho_num}
    \frac{d}{d\delta} \Pr(c_i \in \mathcal{R}_i) = \sum_{j \in \{L, M\}} p_j \, q_j'(\delta) \big[ \Pr(c_i \in \mathcal{R}_i \mid j_i^\star = j, \mathcal{R}_i = \{j\}) - \Pr(c_i \in \mathcal{R}_i \mid j_i^\star = j, \mathcal{R}_i = \{S\}) \big].
\end{align}

We compare the two derivatives term by term. Conditional on $\{j_i^\star = j\}$ and $\mathcal{R}_i = \{j\}$, the user either consumes the recommended title $j$ or takes the outside option, thus the events $\{c_i \in \mathcal{R}_i\}$ and $\{c_i \neq 0\}$ coincide:
\begin{align*}
    \Pr(c_i \in \mathcal{R}_i \mid j_i^\star = j, \mathcal{R}_i = \{j\}) = \Pr(c_i \neq 0 \mid j_i^\star = j, \mathcal{R}_i = \{j\}).
\end{align*}
Conditional on $j_i^\star = j$ and $\mathcal{R}_i = \{S\}$, the user may consume $S$ (the recommended title) or some other title outside the recommendation, thus consuming the recommendation is weakly less likely than consuming anything:
\begin{align*}
    \Pr(c_i \in \mathcal{R}_i \mid j_i^\star = j, \mathcal{R}_i = \{S\}) \leq \Pr(c_i \neq 0 \mid j_i^\star = j, \mathcal{R}_i = \{S\}).
\end{align*}
Subtracting the second from the first, the bracketed term in \eqref{eq:rho_num} weakly exceeds the bracketed term in \eqref{eq:rho_den}. Since the weights $p_j q_j'(\delta)$ are nonnegative with $p_M q_M'(\delta) > 0$,
\begin{align*}
    \frac{d}{d\delta} \Pr(c_i \in \mathcal{R}_i) \geq E'(\delta) > 0.
\end{align*}
Applying the quotient rule, we can write
\begin{align*}
    \rho'(\delta) = \frac{\frac{d}{d\delta} \Pr(c_i \in \mathcal{R}_i) \cdot E(\delta) - \Pr(c_i \in \mathcal{R}_i) \cdot E'(\delta)}{E(\delta)^2}.
\end{align*}
Substituting $\frac{d}{d\delta} \Pr(c_i \in \mathcal{R}_i) \geq E'(\delta)$ and using $E'(\delta) > 0$ from Prediction~\ref{prop:engagement}, we can write
\begin{align*}
    \rho'(\delta) \geq \frac{E'(\delta) \cdot E(\delta) - \Pr(c_i \in \mathcal{R}_i) \cdot E'(\delta)}{E(\delta)^2} = \frac{E'(\delta) \big[ E(\delta) - \Pr(c_i \in \mathcal{R}_i) \big]}{E(\delta)^2}.
\end{align*}
We finally show $E(\delta) > \Pr(c_i \in \mathcal{R}_i)$, which gives the strict inequality. Consider users with $j_i^\star = M$ and $\mathcal{R}_i = \{S\}$. By the bounded, strictly positive density of $\theta$, there is a positive-measure set of such users with $U_{iM} > \max\{U_{iS} + b, U_{iL}, 0\}$. On this event, the user consumes $M$ — outside the singleton recommendation $\{S\}$ — hence $c_i \neq 0$ while $c_i \notin \mathcal{R}_i$. This event occurs with positive probability whenever $q_M(\delta) < 1$, thus
\begin{align}\label{equation:1231255333}
    E(\delta) > \Pr(c_i \in \mathcal{R}_i).
\end{align}
Combined with $E'(\delta) > 0$, \eqref{equation:1231255333} yields $\rho'(\delta) > 0$, and we can conclude that $\Pr(c_i \in \mathcal{R}_i \mid c_i \neq 0)$ is strictly increasing in $\delta$.
\end{proof}

\paragraph{Formal condition for Prediction~\ref{prop:match_quality}.} For any $j \in \{L, M\}$, define the event
\begin{equation}
\label{eq:match_quality_switch_event}
\mathcal{S}_j \coloneqq \left\{ U_{iS} > - b,\; 0 < U_{ij} - U_{iS} < b \right\} .
\end{equation}
Conditional on $\{j_i^\star = j\}$, the event $\mathcal{S}_j$ identifies users who
undergo an inframarginal consumption switch when the recommendation is
corrected from $S$ to $j$. The condition $U_{ij}-U_{iS}<b$ means that the recommendation bonus on $S$
overturns the user's true-utility ranking when $S$ is recommended. The
condition $U_{iS}>-b$ ensures that the user consumes rather than taking the
outside option.

For any $j \in \{L, M\}$, define the event
\begin{equation}\label{eq:match_quality_entry_event}
\mathcal{E}_j \coloneqq \left\{
U_{iS} < - b < U_{ij} < 0 \right\} .
\end{equation}
On this event, $U_{ij}<0$ and $U_{iS}+b<0$, hence the user takes the outside option when $S$ is recommended. However, when $j$ is recommended, $U_{ij} + b > 0$, hence the user consumes $j$.

We now introduce
\begin{align}
r_{\mathrm{inf}}(\delta) & \label{eq:match_quality_r_inf} \coloneqq \sum_{j\in\{L,M \}}
p_j q_j'(\delta)
\Pr(\mathcal{S}_j\mid j_i^\star=j) , \\
g_{\mathrm{inf}}(\delta) & \label{eq:match_quality_g_inf} \coloneqq \frac{\sum_{j\in\{L,M\}}
p_jq_j'(\delta) \mathbb{E}\left[
(U_{ij}-U_{iS}) \mathbbm{1} \{\mathcal{S}_j\} \mid j_i^\star=j \right]}{r_{\mathrm{inf}}(\delta)
} , \\
\label{eq:match_quality_u_ext}
u_{\mathrm{ext}}(\delta) & \coloneqq
\frac{\sum_{j \in \{L,M\}}
p_jq_j'(\delta)
\mathbb{E}\left[U_{ij} \mathbbm{1} \{\mathcal E_j\}\mid j_i^\star = j
\right]}{E'(\delta)}.
\end{align}
Prediction~\ref{prop:engagement} establishes $E'(\delta) > 0$, and we show in the proof that $r_{\mathrm{inf}}(\delta) >0$, hence the ratios are well-defined. We refer to $r_{\mathrm{inf}}(\delta)$ as the rate of inframarginal switching (from $S$ to the user's best title), to $g_{\mathrm{inf}}(\delta)$ as the expected true-utility gain conditional on an inframarginal switch, and to
$u_{\mathrm{ext}}(\delta)$ as the expected true utility conditional on
extensive-margin entry. These expectations aggregate across best-title realizations using the marginal precision gains $p_jq_j'(\delta)$. These definitions weight best-title realizations by the same marginal
precision gains $p_jq_j'(\delta)$ that determine which recommendations are
marginally corrected.

Let $Q(\delta)
\coloneqq \mathbb{E}\left[\sum_{k\in\{L,M,S\}}
U_{ik} \mathbbm{1} \{c_i=k\}
\right]$ denote expected realized true utility, assigning zero utility when the user
takes the outside option. Average match quality can then be written as
\begin{align*}
\overline{U}(\delta) = \frac{\mathbb{E} \left[\sum_{k\in\{L,M,S\}}
U_{ik} \mathbbm{1} \{c_i=k\} \right]}{E(\delta)}.
\end{align*}

\paragraph{Prediction~\ref{prop:match_quality}} (Formal). $\overline{U}(\delta)$ is weakly increasing in $\delta$ if and only if
\begin{equation}
\label{eq:match_quality_condition}
\underbrace{
\frac{r_{\mathrm{inf}}(\delta)}
{E'(\delta)}}_{\substack{\text{inframarginal switches} \\
\text{per extensive-margin entry}}}
\;
\underbrace{
g_{\mathrm{inf}}(\delta)
}_{\substack{\text{true-utility gain per} \\
\text{inframarginal switch}}}
\;\geq\;
\underbrace{
\overline U(\delta) - u_{\mathrm{ext}}(\delta)
}_{\substack{\text{match-quality gap per} \\
\text{extensive-margin entry}}}.
\end{equation}
Average match quality is strictly increasing when the inequality is strict,
locally unchanged when it holds with equality, and decreasing when it is
reversed.

\begin{proof}[Proof of Prediction~\ref{prop:match_quality}.]
The overall strategy is to decompose the effect of a marginal increase in
$\delta$ into an inframarginal change in which title the user consumes and an
extensive-margin change in whether the user consumes. We then apply the
quotient rule to the average match quality. Let $c_i(r)$ denote the user's choice when the recommendation is
$\mathcal R_i=\{r\}$.

Fix $j\in\{L,M\}$ and condition on $j_i^\star=j$. Conditional on
$\{j_i^\star = j\}$, we have that $\{ U_{ij}>U_{ik} \text{ for any } k \neq j\}$ almost surely.

When $\mathcal R_i=\{j\}$, the recommendation bonus reinforces the user's true-utility ranking, hence the user chooses either $j$ or the outside option:
\begin{align*}
c_i(j) = \begin{cases}
j, & \text{ if } U_{ij}+b>0,\\
0, & \text{ if } U_{ij}+b<0.
\end{cases}
\end{align*}
When $\mathcal{R}_i=\{S\}$, the recommendation bonus can lift $S$ above $j$, but no third title can be chosen, as shown previously. Therefore
\begin{align*}
c_i(S) = \begin{cases}
S, \text{ if } & U_{iS} + b > \max\{U_{ij}, 0\} , \\
j, & \text{ if } U_{ij} > \max\{U_{iS} + b,0\},\\
0, & \text{ if } 0 > \max\{U_{ij}, U_{iS} + b\}.
\end{cases}
\end{align*}
Conditional on $\mathcal{S}_j$ and $j_i^\star = j$, correcting the recommendation changes which title the user consumes but not whether the user consumes, hence
\begin{align}\label{equation:match34342}
\mathbbm{1}\{c_i(j) \neq 0 \} -\mathbbm{1}\{c_i(S) \neq 0 \} = 0 .
\end{align}
The resulting increase in realized true utility is $U_{ij} -U_{iS} \in (0,b)$. 

Conditional on $\mathcal{E}_j$ and $j_i^\star = j$, correcting the recommendation raises engagement by one, following 
\begin{align}\label{eq:match_quality_entry_event_bis}
\mathbbm{1}\{c_i(j) \neq 0 \} -\mathbbm{1}\{c_i(S) \neq 0 \} = 1 ,
\end{align}
and changes the realized true utility by $U_{ij} \in (-b, 0)$ (decrease).

In all other preference realizations, correcting the recommendation leaves
both engagement and realized true utility unchanged. Therefore, it follows pointwise from \eqref{equation:match34342} and \eqref{eq:match_quality_entry_event_bis} that
\begin{equation}
\label{eq:match_quality_engagement_identity}
\mathbbm{1}\{c_i(j) \neq 0 \} -\mathbbm{1}\{c_i(S) \neq 0 \} = \mathbbm{1} \{\mathcal E_j\} ,
\end{equation}
and we refer to $\mathbbm{1}\{\mathcal E_j\}$ as the \emph{extensive-margin entry}. The change in utility can be written
\begin{equation}
\label{eq:match_quality_utility_identity}
\sum_{k\in\{L,M,S\}} U_{ik} \mathbbm{1}\{c_i(j)=k\} - \sum_{k\in\{L,M,S\}} U_{ik}\mathbbm{1}\{c_i(S) = k\}  = \underbrace{
(U_{ij}-U_{iS}) \mathbbm{1} \{\mathcal{S}_j\}
}_{\substack{\text{true-utility gain from} \\
\text{an inframarginal switch}}} + \underbrace{U_{ij} \mathbbm{1} \{\mathcal E_j\}
}_{\substack{\text{true utility loss from} \\
\text{extensive-margin entry}}}.
\end{equation}
We now aggregate across the preference realizations whose recommendations are
marginally changed by an increase in $\delta$. From \eqref{eq:E_deriv} in the proof of Prediction~\ref{prop:engagement}, we can write
\begin{align*}
    E'(\delta) = \sum_{j \in \{L, M\}} p_j \, q_j'(\delta) \big[ \Pr(c_i \neq 0 \mid j_i^\star = j, \mathcal{R}_i = \{j\}) - \Pr(c_i \neq 0 \mid j_i^\star = j, \mathcal{R}_i = \{S\}) \big].
\end{align*}
The difference in engagement probabilities is exactly the probability that correcting the recommendation induces extensive-margin entry, hence
\begin{align*}
\Pr(c_i\neq 0 \mid j_i^\star = j, \mathcal R_i = \{j\}) - \Pr(c_i\neq 0 \mid j_i^\star=j, \mathcal R_i=\{S\})
& = \Pr(U_{ij}+b>0, U_{ij}<0, U_{iS} + b < 0 \mid j_i^\star= j ) \\
& = \Pr(U_{ij}>-b, U_{ij}<0, U_{iS}<-b \mid j_i^\star= j) \\
& = \Pr(U_{iS}<-b<U_{ij} < 0 \mid j_i^\star= j) \\
& = \Pr(\mathcal{E}_j \mid j_i^\star= j) ,
\end{align*}
and we obtain that
\begin{align}\label{eq:match_quality_E_derivative}
E'(\delta) = \underbrace{
\sum_{j\in\{L,M\}}
p_jq_j'(\delta)
\Pr(\mathcal E_j\mid j_i^\star=j)
}_{\text{rate of extensive-margin entry}} .
\end{align}
By the strictly positive density of the utility shocks, the event $\{
j_i^\star = M, U_{iS} > - b,\; 0<U_{iM}-U_{iS} < b\} = \{j_i^\star = M, \mathcal{S}_M\}$ has strictly positive probability. Since $p_M q_M'(\delta)>0$ by Assumption~\ref{a:tech} (middle-tail dominance), it follows that
$r_{\mathrm{inf}}(\delta) \geq p_M q_M'(\delta) \Pr(\mathcal{S}_M \mid j_i^\star = M) > 0$, where $r_{\mathrm{inf}}(\delta)$ is defined in \eqref{eq:match_quality_r_inf}.

Conditional on $\{\mathcal{S}_j, j_i^\star = j \}, 0< U_{ij} - U_{iS} <b$, hence
\begin{align*}
    0 < \frac{\sum_{j\in\{L,M\}}
p_jq_j'(\delta) \mathbb{E}\left[
(U_{ij}-U_{iS}) \mathbbm{1} \{\mathcal{S}_j\} \mid j_i^\star=j \right]}{r_{\mathrm{inf}}(\delta)
} & < \frac{\sum_{j\in\{L,M\}}
p_jq_j'(\delta) \mathbb{E}\left[
b \mathbbm{1} \{\mathcal{S}_j\} \mid j_i^\star=j \right]}{r_{\mathrm{inf}}(\delta)} \\
& = b \frac{\sum_{j\in\{L,M\}}
p_jq_j'(\delta) \Pr(\mathcal{S}_j \mid j_i^\star = j )}{r_{\mathrm{inf}}(\delta)} \\
& = b ,
\end{align*}
which implies that $0 < g_{\mathrm{inf}}(\delta) < b$. Conditional on $\{\mathcal{E}_j, j_i^\star = j \}, U_{ij} \in (-b, 0)$ and the same algebra as above gives that $- b < u_{\mathrm{ext}}(\delta) < 0$.

Conditional on $j_i^\star = j\in\{L,M\}$, the recommendation is $j$ with
probability $q_j(\delta)$ and $S$ with probability $1-q_j(\delta)$. Hence,
\begin{align}
Q(\delta) \nonumber & = p_S \mathbb{E}\Bigg[\sum_{k\in\{L,M,S\}}
U_{ik}\mathbbm{1}\{c_i(S)=k\}
\mid j_i^\star=S \Bigg] +
\sum_{j\in\{L, M\}}
p_j q_j(\delta) \mathbb{E} \Bigg[\sum_{k\in\{L,M,S\}} U_{ik}\mathbbm{1}\{c_i(j)=k\} \mid j_i^\star=j \Bigg] \\
& \quad +
\sum_{j\in\{L,M\}}
p_j\bigl(1-q_j(\delta)\bigr)
\mathbb{E}\Bigg[
\sum_{k\in\{L,M,S\}}
U_{ik}\mathbbm{1}\{c_i(S)=k\}
\mid j_i^\star=j
\Bigg].
\end{align}
The first term does not depend on $\delta$, and in the remaining terms, only $q_j(\delta)$ depends on $\delta$. Therefore,
\begin{align}
Q'(\delta) & = 
\sum_{j\in\{L,M\}} p_j q_j'(\delta)
\mathbb{E}\Bigg[\sum_{k\in\{L,M,S\}}
U_{ik}\mathbbm{1} \{c_i(j)=k\}
- \sum_{k\in\{L,M,S\}} U_{ik} \mathbbm{1}\{c_i(S)=k\}
\mid j_i^\star=j \Bigg] \\
& = \sum_{j\in\{L,M\}} p_j q_j'(\delta)
\mathbb{E}\Bigg[(U_{ij} - U_{iS})\mathbbm{1}\{\mathcal{S}_j\} + U_{ij} \mathbbm{1}\{\mathcal{E}_j\} \mid j_i^\star = j\Bigg] \\
& \label{eq:match_quality_Q_derivative} = \underbrace{
r_{\mathrm{inf}}(\delta)g_{\mathrm{inf}}(\delta)
}_{\substack{\text{true-utility gains from} \\
\text{inframarginal switches}}}
+ \underbrace{
E'(\delta) u_{\mathrm{ext}}(\delta)
}_{\substack{\text{true utility contributed by}\\
\text{extensive-margin entry}}} ,
\end{align}
where we use \eqref{eq:match_quality_utility_identity} in the penultimate equality, and the definition of $r_{\mathrm{inf}}(\delta), g_{\mathrm{inf}}(\delta), u_{\mathrm{ext}}(\delta)$ in the last equality.

Since we can write $\overline{U}(\delta) = Q(\delta) / E(\delta)$, applying the quotient rule gives
\begin{align*}
\overline{U}'(\delta) = \frac{
Q'(\delta) E(\delta)-Q(\delta)E'(\delta)}{E(\delta)^2 } =
\frac{1}{E(\delta)}
\left[Q'(\delta)-\overline U(\delta)E'(\delta) \right] = \frac{E'(\delta)}{E(\delta)}
\left[\frac{Q'(\delta)}{E'(\delta)} -\overline U(\delta) \right] .
\end{align*}
Substituting \eqref{eq:match_quality_Q_derivative}, we obtain
\begin{align}\label{eq:match_quality_final_derivative}
\overline U'(\delta) = \frac{E'(\delta)}{E(\delta)}
\Bigg[& \underbrace{
\frac{r_{\mathrm{inf}}(\delta)}{E'(\delta)} g_{\mathrm{inf}}(\delta)
}_{\substack{\text{inframarginal true-utility gains} \\
\text{per extensive-margin entry}}} - \underbrace{\left(\overline U(\delta)-u_{\mathrm{ext}}(\delta)\right)
}_{\substack{\text{match-quality gap per} \\
\text{extensive-margin entry}}}
\Bigg].
\end{align}
Since $E(\delta)>0$ and $E'(\delta)>0$, the sign of
$\overline U'(\delta)$ is exactly the sign of the difference between the two
underbraced terms in
\eqref{eq:match_quality_final_derivative}. Therefore, average match quality is
weakly increasing if and only if
\eqref{eq:match_quality_condition} holds. It is strictly increasing when the
inequality is strict, locally unchanged when it holds with equality, and
decreasing when it is reversed.
\end{proof}

\section{Bayesian Recommender System Microfoundation}\label{sec:recsys_microfoundation}
\begin{figure}[H]
\caption{Recommendation Technology and Matching by Title Bucket}\label{fig:bayes_match}
\begin{center}
\begin{tikzpicture}[x=1.4cm,y=1.8cm]
  \fill[myblue!8] (1.6,0) rectangle (2.8,3.05);
  \node[font=\small,text=myblue!70!black,align=center] at (2.2,-0.3) {empirically relevant range};
  \draw[-latex,black!60] (0,0) -- (8.6,0) node[below left,font=\small] {technology $\delta$};
  \draw[-latex,black!60] (0,0) -- (0,3.3) node[above right,font=\small] {$p_j\,q_j(\delta)$: share of all users matched to $j$};
  \draw[myred,thick] plot[domain=0:8,samples=80] (\x,{3*0.55*(1-0.25*exp(-1.5*\x))});
  \node[font=\small,text=myred,anchor=west] at (8.05,1.65) {$S$};
  \draw[mygreen!60!black,thick] plot[domain=0:8,samples=80] (\x,{3*0.32*(1-exp(-0.55*\x))});
  \node[font=\small,text=mygreen!60!black,anchor=west] at (8.05,0.95) {$M$};
  \draw[black!60,thick] plot[domain=0:8,samples=80] (\x,{3*0.13*(1-exp(-0.12*\x))});
  \node[font=\small,text=black!60,anchor=west] at (8.05,0.25) {$L$};
\end{tikzpicture}
\end{center}
\caption*{\footnotesize \textsc{Notes}: Illustrative curves for the share of users whose best title is $j$ and who are correctly matched to it, $p_j q_j(\delta)$, as technology $\delta$ improves. The shaded region marks the empirically relevant range of $\delta$, where the marginal gain is largest for $M$.}
\end{figure}
In this section we describe a simple Bayesian RecSys that microfounds the two key components of Assumption~\ref{a:tech}: the popularity default and the larger marginal gains for middle-tail titles. Assumption~\ref{a:tech} directly imposes these, but in this section we show that they can arise endogenously from a RecSys that forms beliefs about user preferences under noisy signals, where less popular titles are observed less precisely.

The Bayesian recommender system directly observes the vertical component of preferences ($\mu_j$) and a noisy, unbiased signal about the horizontal component of preferences ($\theta_{ij}$). For expositional ease we impose a Gaussian assumption on $\theta_{ij}$ such that $\theta_{ij} \sim \mathcal{N}(0, \sigma^{2}_{\theta})$ and the platform observes a noisy signal $\eta_{ij} = \theta_{ij} + \epsilon_{ij}/\sqrt{\alpha_j}$ for $j \in \{L, M, S\}$, with $\epsilon_{ij} \sim \mathcal{N}(0,1)$ independent across $(i,j)$ and across $\theta$ so that $\eta_{ij} \mid \theta_{ij} \sim \mathcal{N}(\theta_{ij}, 1/\alpha_j)$. The RecSys holds a correctly specified prior over the unobserved $\theta_{ij}$ -- namely its true marginal $\mathcal{N}(0, \sigma^{2}_{\theta})$.

The key assumption on the recommendation technology is that the signal precision is given by $\alpha_j = \delta \lambda_j$, where $\delta$ indexes technology quality as in our primary model and $\lambda_L < \lambda_M < \lambda_S$ captures the fact that more popular titles have more interaction data. The latter component captures the popularity bias that has been widely documented and studied in the RecSys literature \citep{celma2008hits, klimashevskaia2024survey}.

Since $\mu_j$ is known and the prior on $\theta_{ij}$ is conjugate to the Gaussian signal, the posterior mean of utility $U_{ij} = \mu_j + \theta_{ij}$ is:
\begin{align*}
s_{ij} \;=\; \mu_j + w(\alpha_j)\,\eta_{ij},
\qquad w(\alpha) \;=\; \frac{\alpha\sigma^{2}_{\theta}}{1 + \alpha\sigma^{2}_{\theta}},
\qquad j \in \{L, M, S\},
\end{align*}
where the shrinkage weight $w(\alpha_j)$ is the posterior precision ratio from the Gaussian Bayesian update. The recommender scores each title by this posterior mean and
recommends $\mathcal{R}_i = \{\arg\max_{j \in \{L,M,S\}} s_{ij}\}$.

This structure motivates each component of Assumption~\ref{a:tech}:
\begin{enumerate}
    \item[(i)] \textbf{Popularity default.} In the limit case when the recommendation technology is very poor (i.e., $\delta$ close to $0$), $w(\alpha_j) \to 0$ for all $j \in \{L, M, S\}$. In this case, the score assigned by the RecSys to each title collapses to the vertical component ($s_{ij} \to \mu_j$) and consequently the recommender selects $S$ for every user as $\mu_S > \mu_M > \mu_L$. Thus, in this limit case only the superstar title can be recommended to users.

As $\delta$ increases, the RecSys puts more weight on the horizontal component and decreases the degree of this popularity default but does not remove it entirely. The consequence is that a less popular title $j$ is recommended over $S$ only when its weighted signal overcomes the popularity advantage of $S$, namely when $w(\alpha_j)\,\eta_{ij} - w(\alpha_S)\,\eta_{iS} > \mu_S - \mu_j$. Since $w(\alpha_j)$ is increasing in $\delta$, the RecSys correctly recommends $j$ to a larger share of the users who prefer it as the technology improves. The popularity
default nonetheless persists at every finite $\delta$ -- a user is still recommended $S$ whenever the signal for their preferred title is insufficiently favorable -- but this bias weakens as the technology improves. This is the stylized ``popularity default" of Assumption~\ref{a:tech}: the recommender defaults to $S$ whenever the signal for a user's preferred title is insufficiently favorable, which weakens as $\delta$ increases but always persists unless the technology is perfect.

    \item[(ii)] \textbf{$q_j$ monotonicity in $\delta$.} The weight $w(\alpha_j)$ is strictly increasing in $\delta$ for each $j \in \{L, M, S\}$, so the score puts increasing weight on the horizontal component $\theta_{ij}$ relative to the vertical component $\mu_j$. As a result, the probability that the recommender identifies the user's best title, $q_j(\delta)$, is weakly increasing in $\delta$ for each $j$. For $S$, this monotonicity may be nearly flat as $q_S$ is precise at realistic values for $\delta$.

   \item[(iii)] \textbf{Middle-tail dominance.} Assumption~\ref{a:tech} requires the marginal gain in correctly matched users, $p_j q_j'(\delta)$ -- the marginal precision gain $q_j'(\delta)$ weighted by the share of users $p_j$ for whom $j$ is the preferred title -- to be larger for the middle-tail $M$ than for the long-tail $L$ in the empirically relevant range of $\delta$.

This follows since the long-tail $L$ has low baseline popularity $\mu_L$ -- a large gap between it and the superstar title -- and low signal precision $\lambda_L$ -- a direct consequence of $L$ generating less interaction data. Together these cause $L$'s effective precision $\alpha_L = \delta \lambda_L$ to advance slowly, so at empirically relevant $\delta$ its noisy signal remains far from the threshold to switch the recommendation from $S$ to $L$, and $q'_L(\delta)$ stays modest. The middle-tail $M$, on the other hand, is closer to that threshold to begin with and, because $\lambda_M > \lambda_L$, crosses it faster as $\delta$ improves. Together these lead to
\begin{align*}
    p_M q_M'(\delta) > p_L q_L'(\delta),
\end{align*}
as long as $L$'s marginal precision gain does not outweigh $M$'s larger share of users for whom it is their preferred title. 

We provide an illustration of this in Figure~\ref{fig:bayes_match} whereby $S$ benefits the most for small $\delta$, $M$ for most intermediate values of $\delta$ until it flattens, and then $L$ as $\delta$ gets arbitrarily large. Thus, middle-tail dominance need not hold for all $\delta$ -- at very high $\delta$, $M$ saturates while $L$ continues to advance -- but should hold for the relevant portion of $\delta$ when the RecSys is not already matching $M$ users to their preferred title and is matching $S$ users to their preferred title, which we believe is relevant for a mature RecSys such as the one deployed at Netflix.
\end{enumerate}
\end{document}